\documentclass{llncs}

\usepackage[usenames,dvipsnames]{color}

\usepackage{latexsym}
\usepackage{amsxtra} 
\usepackage{amssymb}
\usepackage{amsmath}
\usepackage{pslatex}
\usepackage{epsfig}
\usepackage{wrapfig}
\usepackage{paralist}
\usepackage{txfonts}
\usepackage{framed}
\usepackage{makecell}
\usepackage{url}
\usepackage{tikz}
\usetikzlibrary{automata,positioning, calc}
\usepackage[inline,shortlabels]{enumitem}

\usepackage{proof}

\usepackage{algorithm}
\usepackage{algorithmicx}
\usepackage[noend]{algpseudocode}

\usepackage[draft]{commenting}

\newif\ifLongVersion\LongVersiontrue

\newtheorem{fact}{Fact}

\newcommand{\rbr}{{\bf ]\!]}}
\newcommand{\lbr}{{\bf [\![}}
\newcommand{\sem}[1]{\lbr #1 \rbr}

\newcommand{\set}[1]{\{ #1 \}}
\newcommand{\tuple}[1]{\langle #1 \rangle}
\renewcommand{\vec}[1]{\mathbf #1}

\newcommand{\minsem}[1]{\sem{#1}^\mu}

\newcommand{\isdef}{\stackrel{\scriptscriptstyle{\mathsf{def}}}{=}}
\renewcommand{\iff}{\Leftrightarrow}

\newcommand{\stamp}[2]{{#1}^{\scriptscriptstyle{({#2})}}}
\newcommand{\overstamp}[2]{{#1}^{\scriptscriptstyle{\langle{#2}\rangle}}}

\newcommand{\len}[1]{{|{#1}|}}

\newcommand{\lang}[1]{{\mathcal L}({#1})}
\newcommand{\clang}[1]{{\mathcal L}^c({#1})}

\newcommand{\arrow}[2]{\xrightarrow{{\scriptscriptstyle #1}}_{{\scriptstyle #2}}}

\newcommand{\nat}{{\bf \mathbb{N}}}
\newcommand{\zed}{{\bf \mathbb{Z}}}
\newcommand{\rat}{{\bf \mathbb{Q}}}
\newcommand{\real}{{\bf \mathbb{R}}}

\renewcommand{\paragraph}[1]{\noindent{\bf #1}}

\renewcommand{\proof}[1]{\ifLongVersion \noindent\emph{Proof}: {#1} \vspace*{\baselineskip}\else\fi}

\newcommand{\teq}{\approx}

\newcommand{\I}{\mathcal{I}}
\newcommand{\J}{\mathcal{J}}
\newcommand{\G}{\mathcal{G}}
\newcommand{\K}{\mathcal{K}}

\newcommand{\fv}[2]{\mathrm{FV}_{#1}(#2)}
\newcommand{\pset}[2]{\mathrm{P}^{#1}(#2)}
\newcommand{\voc}[1]{\mathrm{V}({#1})}

\newcommand{\dom}{\mathrm{dom}}

\newcommand{\vars}{\mathsf{Var}}
\newcommand{\preds}{\mathsf{Pred}}

\newcommand{\Data}{\mathbb{D}}
\newcommand{\Bool}{\mathbb{B}}

\newcommand{\dual}[1]{{#1}^{\sim}}
\newcommand{\bigO}[1]{\mathcal{O}({#1})}

\newcommand{\A}{\mathcal{A}}

\newcommand{\impact}{IMPACT}

\newcommand{\cube}[1]{\mathsf{c}({#1})}
\newcommand{\T}{\mathcal{T}}
\newcommand{\event}[1]{{#1}_\Sigma}
\newcommand{\data}[1]{{#1}_\Data}
\newcommand{\pathform}[1]{\Theta({#1})}
\newcommand{\accform}[1]{\Upsilon({#1})}
\newcommand{\accformA}[2]{\Upsilon_{#1}({#2})}
\newcommand{\substform}[1]{\overline{\Upsilon}({#1})}
\newcommand{\substformn}[1]{\overline{\Theta}({#1})}
\newcommand{\substformA}[2]{\overline{\Upsilon}_{#1}({#2})}        
\newcommand{\quantform}[1]{\widehat{\Upsilon}({#1})}
\newcommand{\quantformn}[1]{\widehat{\Theta}({#1})}

\newcommand{\prefix}{\preceq}
\newcommand{\posforms}{\mathsf{Form}^+}
\newcommand{\cover}{\sqsubseteq}

\declareauthor{ri}{Radu}{blue}
\declareauthor{xx}{Xiao}{red}

\begin{document}

\title{First Order Alternation}

\author{Radu Iosif and Xiao Xu}
\institute{
  CNRS, Verimag, Universit\'e de Grenoble Alpes \\
  Email: \{Radu.Iosif,Xiao.Xu\}@univ-grenoble-alpes.fr
}

\maketitle

\begin{abstract}
We introduce first order alternating automata, a generalization of
boolean alternating automata, in which transition rules are described
by multisorted first order formulae, with states and internal
variables given by uninterpreted predicate terms. The model is closed
under union, intersection and complement, and its emptiness problem is
undecidable, even for the simplest data theory of equality. To cope
with this limitation, we develop an abstraction refinement
semi-algorithm based on lazy annotation of the symbolic execution
paths with interpolants, obtained by applying (i) quantifier
elimination with witness term generation and (ii) Lyndon interpolation
in the quantifier-free data theory with uninterpreted predicate
symbols. This provides a method for checking inclusion of timed and
finite-memory register automata, and emptiness of quantified predicate
automata, previously used in the verification of parameterized
concurrent programs, composed of replicated threads, with a
shared-memory communication model.
\end{abstract}

\section{Introduction}

Many results in formal language theory rely on the assumption that
languages are defined over finite alphabets. In practice, this
assumption is problematic when attempting to use automata as models of
real-time systems or even simple programs, whose input and observable
output requires taking into account data values, ranging over very
large domains, better viewed as infinite mathematical abstractions.

Alternating automata are a generalization of nondeterministic automata
with universal transitions, that create several copies of the
automaton, which synchronize on the same input word. Alternating
automata are appealing for verification because they allow encoding of
problems such as temporal logic model checking in linear time, as
opposed to the exponential time required by nondeterministic automata
\cite{Vardi95}. A finite-alphabet alternating automaton is typically
described by a set of transition rules $q \arrow{a}{} \phi$, where $q$
is a state, $a$ is an input symbol and $\phi$ is a positive boolean
combinations of states, viewed as propositional variables.

Here we introduce a generalized alternating automata model in which
states are predicate symbols $q(y_1,\ldots,y_k)$, the input has
associated data variables $x_1,\ldots,x_n$, ranging over an infinite
domain and transitions are of the form $q(y_1,\ldots,y_k)
\arrow{a(x_1,\ldots,x_n)}{} \phi$, where $\phi$ is any formula in the
first-order theory of the data domain, in which each state predicate
occurs under an even number of negations. In this model, the arguments
of a predicate atom $q(y_1,\ldots,y_k)$ track the values of the
\emph{internal variables} associated with the state. Together with the
input values $x_1,\ldots,x_n$, these values are used to compute the
successor states and are invisible in the input sequence.

Previous attempts to generalize classical Rabin-Scott automata to
infinite alphabets, such as timed automata \cite{AlurDill94} and
finite-memory (register) automata \cite{KaminskiFrancez94} face the
\emph{complement closure} problem: there exist automata for which the
complement language cannot be recognized by an automaton in the same
class. This excludes the possibility of encoding a language inclusion
problem $\lang{A} \subseteq \lang{B}$ as the emptiness of an automaton
recognizing the language $\lang{A} \cap \clang{B}$, where $\clang{B}$
denotes the complement of $\lang{B}$. 

The solution we adopt here is a tight coupling of internal variables
to control states, using uninterpreted predicate symbols. As we show,
this allows for linear-time complementation just as in the case of
boolean alternating automata. Complementation is, moreover, possible
when the transition formulae contain first-order quantifiers,
generating infinitely-branching execution trees. The price to be paid
for this expressivity is that emptiness of first-order alternating
automata is undecidable, even for the simplest data theory of equality
\cite{Farzan15}.

The main contribution of this paper is an effective emptiness checking
semi-algorithm for first-order alternating automata, in the spirit of
the \impact~procedure, originally developed for checking safety of
nondeterministic integer programs \cite{McMillan06}. However, checking
emptiness of first-order alternating automata by lazy annotation with
interpolants faces two problems: \begin{compactenum}
\item Quantified transition rules make it hard, or even impossible, to
  decide if a given symbolic trace is spurious. This is mainly because
  adding uninterpreted predicate symbols to decidable first-order
  theories, such as Presburger arithmetic, results in undecidability
  \cite{Halpern91}. To deal with this problem, we assume that the
  first order data theory, without uninterpreted predicate symbols,
  has a quantifier elimination procedure, that instantiates quantifers
  with effectively computable \emph{witness terms}.
\item The interpolants that prove the spuriousness of a symbolic path
  are not \emph{local}, as they may refer to input values encountered
  in the past. However, the future executions are oblivious to
  \emph{when} these values have been seen in the past and depend only
  on the data constraints between the values. We use this fact to
  define a labeling of nodes, visited by the lazy annotation
  procedure, with conjunctions of existentially quantified
  interpolants combining predicate atoms with data constraints.
\end{compactenum}

As applications of first order alternating automata, we identified
several undecidable problems for which no semi-algorithmic methods
exist: inclusion between recognizable timed languages
\cite{AlurDill94}, languages recognized by finite-memory automata
\cite{KaminskiFrancez94} and emptiness of predicate automata, a
subclass of first-order alternating automata used to check safety and
liveness properties of parameterized concurrent programs
\cite{Farzan15,Farzan16}.

For reasons of space, all proofs of technical results in this paper
are given in \cite{IosifXu18ArXiv}.

\paragraph{\bf Related Work} 
The first order alternating automata model presented in this paper
stems from our previous work on boolean alternating automata extended
with variables ranging over infinite data \cite{IosifXu18}. There we
considered states to be propositional variables, as in the classical
textbook alternating automata model, and all variables of the
automaton to be observable in the input. The model in this paper
overcomes this latter restriction by allowing for internal variables,
whose variables are not visible in the language. 

This solves an older language inclusion problem
$\bigcap_{i=1}^n\lang{A_i} \subseteq \lang{B}$, between finite-state
automata with data variables, whose languages are alternating
sequences of input events and variable valuations
\cite{IosifRV16}. There, we assumed that all variables of the observer
automaton $B$ must be declared in the automata $A_1, \ldots, A_n$ that
model the concurrent components of the system under check. Using
first-order alternating automata allows to bypass this limitation of
our previous work.

The work probably closest to the one reported here concerns the model
of predicate automata (PA) \cite{Farzan15,Farzan16,KincaidPhD},
applied to the verification of parameterized concurrent programs with
shared memory. In this model, the alphabet consists of pairs of
program statements and thread identifiers, thus being infinite because
the number of threads is unbounded. Because thread identifiers can
only be compared for (dis-)equality, the data theory in PA is the
theory of equality. Even with this simplification, the emptiness
problem is undecidable when either the predicates have arity greater
than one \cite{Farzan15} or quantified transition rules
\cite{KincaidPhD}. Checking emptiness of quantifier free PA is
possible semi-algorithmically, by explicitly enumerating reachable
configurations and checking coverage by looking for permuations of
argument values. However, no semi-algorithm is given for quantified
PA.  Dealing with quantified transition rules is one of the
contributions of the work reported in this paper. 

\section{Preliminaries}

For two integers $0 \leq i \leq j$, we denote by $[i,j]$ the set
$\set{i,i+1,\ldots,j}$ and by $[i]$ the set $[0,i]$. We consider two
sorts $\Data$ and $\Bool$, where $\Data$ is an infinite domain and
$\Bool = \set{\top,\bot}$ is the set of boolean values true ($\top$)
and false ($\bot$), respectively. The $\Data$ sort is equipped with
finitely many function symbols $f : \Data^{\#(f)} \rightarrow \Data$,
where $\#(f)\geq0$ denotes the number of arguments (arity) of
$f$. When $\#(f)=0$, we say that $f$ is a constant. A \emph{predicate}
is a function symbol $p : \Data^{\#(p)} \rightarrow \Bool$, denoting a
relation of arity $\#(p)$ and we write $\preds$ for the set of
predicates. 

In the following, we shall consider that the interpretation of all
function symbols $f : \Data^{\#(f)} \rightarrow \Data$ that are not
predicates is fixed by the interpretation of the $\Data$ sort,
e.g.\ if $\Data$ is the set of integers $\zed$, the function symbols
are zero, the successor function and the arithmetic operations of
addition and multiplication. For simplicity, we further blur the
notational distinction between function symbols and their
interpretations.

Let $\vars = \set{x,y,z,\ldots}$ be an infinite countable set of
variables, ranging over $\Data$. Terms are either constants of sort
$\Data$, variables or function applications $f(t_1,\ldots,t_{\#(f)})$,
where $t_1,\ldots,t_{\#(f)}$ are terms. The set of first order
formulae is defined by the syntax below:
\[\phi := t \teq s \mid p(t_1,\ldots,t_{\#(p)}) 
\mid \neg \phi_1 \mid \phi_1 \wedge \phi_2 \mid \exists x ~.~
\phi_1 \] where $t,s,t_1,\ldots,t_{\#(p)}$ denote terms. We write
$\phi_1 \vee \phi_2$, $\phi_1 \rightarrow \phi_2$ and $\forall x ~.~
\phi_1$ for $\neg(\neg\phi_1 \wedge \neg\phi_2)$, $\neg\phi_1 \vee
\phi_2$ and $\neg\exists x ~.~ \neg\phi_1$, respectively. We denote by
$\fv{}{\phi}$ the set of free variables in $\phi$. The size
$\len{\phi}$ of a formula $\phi$ is the number of symbols needed to
write it down.

A \emph{sentence} is a formula $\phi$ in which each variable occurs
under the scope of a quantifier, i.e. $\fv{}{\phi} = \emptyset$.  A
formula is \emph{positive} if each predicate symbol occurs under an
even number of negations and we denote by $\posforms(Q,X)$ the set of
positive formulae with predicates from the set $Q \subseteq \preds$
and free variables from the set $X \subseteq \vars$.

A formula is in \emph{prenex form} if it is of the form $\varphi =
Q_1x_1 \ldots Q_nx_n ~.~ \phi$, where $\phi$ has no quantifiers. In
this case we call $\phi$ the \emph{matrix} of $\varphi$. Every
first order formula can be written in prenex form, by renaming each
quantified variable to a unique name and moving the quantifiers
upfront.


An \emph{interpretation} $\I$ maps each predicate $p$ into a set $p^\I
\subseteq \Data^{\#(p)}$, if $\#(p)>0$, or into an element of $\Data$
if $\#(p)=0$. A \emph{valuation} $\nu$ maps each variable $x$ into an
element of $\Data$. Given a term $t$, we denote by $t^\nu$ the value
obtained by replacing each variable $x$ by the value $\nu(x)$ and
evaluating each function application. For a formula $\phi$, we define
the forcing relation $\I,\nu \models \phi$ recursively on the
structure of $\phi$, as usual. 
\ifLongVersion
\[
\begin{array}{rcl}
\I,\nu \models t \teq s & \iff & t^\nu=s^\nu \\
\I,\nu \models p(t_1,\ldots,t_{\#(p)}) & \iff & \tuple{t_1^\nu,\ldots,t_{\#(p)}^\nu} \in p^\I \\
\I,\nu \models \neg\phi_1 & \iff & \I,\nu \not\models \phi_1 \\
\I,\nu \models \phi_1 \wedge \phi_2 & \iff & \I,\nu \models \phi_i \text{, for all } i=1,2 \\ 
\I,\nu \models \exists x ~.~ \phi_1 & \iff & \I,\nu[x\leftarrow d] \models \phi_1 \text{, for some } d \in \Data 
\end{array}
\]
where $\nu[x\leftarrow d]$ is the valuation which assigns $d$ to $x$
and behaves like $\nu$ elsewhere.  
\fi 
For a formula $\phi$ and a valuation $\nu$, we define $\sem{\phi}_\nu
\isdef \set{\I \mid \I,\nu \models \phi}$ and drop the $\nu$ subscript
for sentences. A sentence $\phi$ is \emph{satisfiable}
(\emph{unsatisfiable}) if $\sem{\phi}\neq\emptyset$
($\sem{\phi}=\emptyset$). An element of $\sem{\phi}$ is called a
\emph{model} of $\phi$. A formula $\phi$ is \emph{valid} if $\I,\nu
\models \phi$ for every interpretation $\I$ and every valuation
$\nu$. For two formulae $\phi$ and $\psi$ we write $\phi \models \psi$
for $\sem{\phi} \subseteq \sem{\psi}$, in which case we say that
$\phi$ \emph{entails} $\psi$.

Interpretations are partially ordered by the pointwise subset order,
defined as $\I_1 \subseteq \I_2$ if and only if $p^{\I_1} \subseteq
p^{\I_2}$ for each predicate $p \in \preds$. Given a set $\mathcal{S}$
of interpretations, a minimal element $\I \in \mathcal{S}$ is an
interpretation such that for no other interpretation $\I' \in
\mathcal{S} \setminus \set{\I}$ do we have $\I' \subseteq \I$. For a
formula $\phi$ and a valuation $\nu$, we denote by $\minsem{\phi}_\nu$
and $\minsem{\phi}$ the set of minimal interpretations from
$\sem{\phi}_\nu$ and $\sem{\phi}$, respectively.

\section{First Order Alternating Automata}

Let $\Sigma$ be a finite alphabet $\Sigma$ of \emph{input
  events}. Given a finite set of variables $X \subseteq \vars$, we
denote by $X \mapsto \Data$ the set of valuations of the variables $X$
and $\Sigma[X] = \Sigma \times (X \mapsto \Data)$ be the possibly
infinite set of \emph{data symbols} $(a,\nu)$, where $a$ is an input
symbol and $\nu$ is a valuation. A \emph{data word} (simply called
word in the following) is a finite sequence $w=(a_1,\nu_1)(a_2,\nu_2)
\ldots (a_n,\nu_n)$ of data symbols. Given a word $w$, we denote by
$\event{w} \isdef a_1\ldots a_n$ its sequence of input events and by
$\data{w}$ the valuation associating each time-stamped variable
$\stamp{x}{i}$ the value $\nu_i(x)$, for all $x \in \vars$ and
$i\in[1,n]$. We denote by $\varepsilon$ the empty sequence, by
$\Sigma^*$ the set of finite sequences of input events and by
$\Sigma[X]^*$ the set of data words over the variables $X$. 

Formally, a \emph{first order alternating automaton} is a tuple $\A =
\tuple{\Sigma,X,Q,\iota,F,\Delta}$, where $\Sigma$ is a finite set of
input events, $X$ is a finite set of input variables, $Q$ is a finite
set of predicates denoting control states, $\iota \in
\posforms(Q,\emptyset)$ is a sentence defining initial configurations,
$F \subseteq Q$ is the set of predicates denoting final states, and
$\Delta$ is a set of \emph{transition rules} of the form
\(q(y_1,\ldots,y_{\#(q)}) \arrow{a(X)}{} \psi\), where $q \in Q$ is a
predicate, $a \in \Sigma$ is an input event and $\psi \in
\posforms(Q,X \cup \set{y_1,\ldots,y_{\#(q)}})$ is a positive formula,
where $X \cap \set{y_1,\ldots,y_{\#(q)}} = \emptyset$. The quantifiers
occurring in the right-hand side formula of a transition rule are
referred to as \emph{transition quantifiers}. The \emph{size} of $\A$
is defined as $\len{\A} = \len{\iota} +
\sum_{\scriptscriptstyle{q(\vec{y}) \arrow{a(X)}{} \psi \in \Delta}}
\len{\psi}$.

The intuition of a transition rule \(q(y_1,\ldots,y_{\#(q)})
\arrow{a(X)}{} \psi\) is the following: $a$ is the input event and $X$
are the input data values that trigger the transition, whereas $q$ and
$y_1,\ldots,y_{\#(q)}$ are the current control state and data values
in that state, respectively. Without loss of generality, we consider,
for each predicate $q \in Q$ and each input event $a \in \Sigma$, at
most one such rule, as two or more rules can be joined using
disjunction.

The execution semantics of automata is given in close analogy with the
case of boolean alternating automata, with transition rules of the
form $q \arrow{a}{} \phi$, where $q$ is a boolean constant and $\phi$
a positive boolean combination of such constants. For instance, $q_0
\arrow{a}{} q_1 \wedge q_2 \vee q_3$ means that the automaton can
choose to transition in either both $q_1$ and $q_2$ or in $q_3$
alone. This intuition leads to saying that the steps of the automaton
are defined by the minimal boolean models of the transition
formulae. In this case, both $\set{q_1 \leftarrow \top, q_2 \leftarrow
  \top, q_3 \leftarrow \bot}$ and $\set{q_1 \leftarrow \bot, q_2
  \leftarrow \bot, q_3 \leftarrow \top}$ are minimal models, however
$\set{q_1 \leftarrow \top, q_2 \leftarrow \top, q_3 \leftarrow \top}$,
is a model but is not minimal. The original definition of alternating
finite-state automata \cite{ChandraKozenStockmeyer81} works around
this problem by considering boolean valuations (models) instead of
formulae. However, describing first-order alternating automata using
interpretations instead of formulae would be rather hard to follow.

Given a predicate $q \in Q$ and a tuple of data values
$d_1,\ldots,d_{\#(q)}$, the tuple $q(d_1,\ldots,d_{\#(q)})$ is called
a \emph{configuration}\footnote{Note that a configuration is not a
  logical term since data values cannot be written in logic.}. To
formalize the execution semantics of automata, we relate sets of
configurations to models of first order sentences, as follows. Each
first-order interpretation $\I$ corresponds to a set of configurations
$\cube{\I} \isdef \set{q(d_1,\ldots,d_{\#(q)}) \mid q \in Q,~
  \tuple{d_1,\ldots,d_{\#(q)}} \in q^\I}$, called a \emph{cube}. For a
set $\mathcal{S}$ of interpretations, we define $\cube{\mathcal{S}}
\isdef \set{\cube{\I} \mid \I \in \mathcal{S}}$.

\begin{definition}\label{def:execution}
Given a word $w=(a_1,\nu_1) \ldots (a_n,\nu_n) \in \Sigma[X]^*$ and a cube $c$, an
\emph{execution} of $\A=\tuple{\Sigma,X,Q,\iota,F,\Delta}$ over $w$, starting with $c$, is
a (possibly infinite) forest $\T = \set{T_1,T_2,\ldots}$, where each
$T_i$ is a tree labeled with configurations, such that:
\begin{compactenum}
\item\label{it1:execution} $c = \set{T(\epsilon) \mid T \in \T}$ is the set of
  configurations labeling the roots of $T_1,T_2,\ldots$ and
\item\label{it2:execution} if $q(d_1,\ldots,d_{\#(q)})$ labels a node on the level $j \in
  [n-1]$ in $T_i$, then the labels of its children form a cube from
  $\cube{\minsem{\psi}_{\eta}}$, where $\eta = \nu_{j+1}[y_1
    \leftarrow d_1,\ldots,y_{\#(q)} \leftarrow d_{\#(q)}]$ and
  \(q(y_1,\ldots,y_{\#(q)}) \arrow{a_{j+1}(X)}{} \psi \in \Delta\) is
  a transition rule of $\A$.
\end{compactenum}
\end{definition}

\begin{definition}\label{def:accepting}
An execution $\T$ over $w$, starting with $c$, is \emph{accepting} if
and only if \begin{compactitem}
\item\label{it1:accepting} all paths in $\T$ have the same length $n$,
  and
\item\label{it2:accepting} the frontier of each tree $T \in \T$ is
  labeled with \emph{final configurations} $q(d_1,\ldots,d_{\#(q)})$,
  where $q \in F$.
\end{compactitem} 
If $\A$ has an accepting execution over $w$ starting with a cube $c
\in \cube{\minsem{\iota}}$, then $\A$ \emph{accepts} $w$ and let
$\lang{\A}$ be the set of words accepted by $\A$.
\end{definition}
In this paper, we address the following questions: \begin{compactenum}
\item \emph{boolean closure}: given automata $\A_i =
  \tuple{\Sigma,X,Q_i,\iota_i,F_i,\Delta_i}$, for $i=1,2$, do there
  exist automata $\A_\cap$, $\A_\cup$ and $\overline{\A}_1$ such that
  $L(\A_\cap) = L(\A_1) \cap L(\A_2)$, $L(\A_\cup) = L(\A_1) \cup
  L(\A_2)$ and $L(\overline{\A}_1) = \Sigma[X]^* \setminus L(\A_1)$ ?
\item \emph{emptiness}: given an automaton $\A$, is $L(\A) =
  \emptyset$?
\end{compactenum}

\subsection{Symbolic Execution}

In the upcoming developments it is sometimes more convenient to work
with logical formulae defining executions of automata, than with
low-level execution forests. For this reason, we first introduce
\emph{path formulae} $\pathform{\alpha}$, which are formulae defining
the executions of an automaton, over words that share a given sequence
$\alpha$ of input events.  Second, we restrict a path formula
$\pathform{\alpha}$ to an \emph{acceptance formula}
$\accform{\alpha}$, which defines only accepting executions over words
that share a given input sequence. Otherwise stated,
$\accform{\alpha}$ is satisfiable if and only if the automaton accepts
a word $w$ such that $\event{w} = \alpha$.

Let $\A = \tuple{\Sigma,X,Q,\iota,F,\Delta}$ be an automaton for the
rest of this section. For any $i \in \nat$, we denote by $\stamp{Q}{i}
= \set{\stamp{q}{i} \mid q \in Q}$ and $\stamp{X}{i} =
\set{\stamp{x}{i} \mid x \in X}$ the sets of time-stamped predicates
and variables, respectively. As a shorthand, we write $\stamp{Q}{\leq
  n}$ (resp.  $\stamp{X}{\leq n}$) for the set $\set{\stamp{q}{i} \mid
  q \in Q, i \in [n]}$ (resp. $\set{\stamp{x}{i} \mid x \in X, i \in
  [n]}$). For a formula $\psi$ and $i \in \nat$, we define
$\stamp{\psi}{i} \isdef \psi[\stamp{X}{i}/X,\stamp{Q}{i}/Q]$ the
formula in which all input variables and state predicates (and only
those symbols) are replaced by their time-stamped counterparts. As a
shorthand, we shall write $q(\vec{y})$ for $q(y_1,\ldots,y_{\#(q)})$,
when no confusion arises.

Given a sequence of input events $\alpha = a_1 \ldots a_n \in
\Sigma^*$, the \emph{path formula} of $\alpha$ is:
\begin{equation}\label{eq:pathform}
\pathform{\alpha} \isdef \stamp{\iota}{0} \wedge 
\bigwedge_{i=1}^n \bigwedge_{q(\vec{y}) \arrow{a_i(X)}{}
  \psi \in \Delta} \forall y_1 \ldots \forall y_{\#(q)} ~.~
\stamp{q}{i-1}(\vec{y}) \rightarrow \stamp{\psi}{i}
\end{equation}
The automaton $\A$, to which $\pathform{\alpha}$ refers, will always
be clear from the context. To formalize the relation between the
low-level configuration-based execution semantics and the symbolic
path formulae, consider a word $w=(a_1,\nu_1) \ldots (a_n,\nu_n) \in
\Sigma[X]^*$. Any execution forest $\T$ of $\A$ over $w$ is associated
an interpretation $\I_{\T}$ of the set of time-stamped predicates
$\stamp{Q}{\leq n}$, defined as:
\[\I_{\T}(\stamp{q}{i}) \isdef \set{\tuple{d_1,\ldots,d_{\#(q)}} \mid
  q(d_1,\ldots,d_{\#(q)}) \text{ labels a node on level $i$ in $\T$}} 
,~\forall q\in Q ~\forall i \in [n]\]

\begin{lemma}\label{lemma:path-formula}
  Given an automaton $\A = \tuple{\Sigma,X,Q,\iota,F,\Delta}$, for any
  word $w=(a_1,\nu_1) \ldots (a_n,\nu_n)$, we have 
  \(\minsem{\pathform{\event{w}}}_{\data{w}} = \set{\I_{\T} \mid \T
    \text{ is an execution of $\A$ over $w$}}\). 
\end{lemma}
\proof{ ``$\subseteq$'' Let $\I$ be a minimal interpretation such that
  $\I,\data{w} \models \pathform{\event{w}}$. We show that there exists
  an execution $\T$ of $\A$ over $w$ such that $\I = \I_\T$, by
  induction on $n\geq 0$. For $n=0$, we have $w=\epsilon$ and
  $\pathform{\event{w}} = \stamp{\iota}{0}$. Because $\iota$ is a
  sentence, the valuation $\data{w}$ is not important in $\I,\data{w}
  \models \stamp{\iota}{0}$ and, moreover, since $\I$ is minimal, we
  have $\I \in \minsem{\stamp{\iota}{0}}$. We define the
  interpretation $\J(q)=\I(\stamp{q}{0})$, for all $q \in Q$. Then
  $\cube{\J}$ is an execution of $\A$ over $\epsilon$ and $\I =
  \I_{\cube{\J}}$ is immediate. For the inductive case $n>0$, we
  assume that $w=u\cdot(a_n,\nu_n)$ for a word $u$. Let $\J$ be the
  interpretation defined as $\I$ for all $\stamp{q}{i}$, with $q \in
  Q$ and $i \in [n-1]$, and $\emptyset$ everywhere else. Then
  $\J,u_\Data \models \pathform{u_\Sigma}$ and $\J$ is moreover
  minimal. By the induction hypothesis, there exists an execution $\G$
  of $\A$ over $u$, such that $\J = \I_\G$. Consider a leaf of a tree
  $T \in \G$, labeled with a configuration $q(d_1,\ldots,d_{\#(q)})$
  and let $\forall y_1 \ldots \forall y_{\#(q)} ~.~
  \stamp{q}{n-1}(\vec{y}) \rightarrow \stamp{\psi}{n}$ be the
  subformula of $\pathform{\event{w}}$ corresponding to the
  application(s) of the transition rule $q(\vec{y}) \arrow{a_n}{}
  \psi$ at the $(n-1)$-th step. Let $\nu = \data{w}[y_1 \leftarrow
    d_1,\ldots,y_{\#(q)} \leftarrow d_{\#(q)}]$. Because $\I, \data{w}
  \models \forall y_1 \ldots \forall y_{\#(q)} ~.~
  \stamp{q}{n-1}(\vec{y}) \rightarrow \stamp{\psi}{n}$, we have $\I
  \in \sem{\stamp{\psi}{n}}_{\nu}$ and let $\K$ be one of the minimal
  interpretations such that $\K \subseteq \I$ and $\K \in
  \sem{\stamp{\psi}{n}}_{\nu}$. It is not hard to see that $\K$ exists
  and is unique, otherwise we could take the pointwise intersection of
  two or more such interpretations. We define the interpretation
  $\overline{\K}(q) = \overline{\K}(\stamp{q}{n})$ for all $q \in
  Q$. We have that $\overline{\K} \in \minsem{\psi}_{\nu}$ --- if
  $\overline{\K}$ was not minimal, $\K$ was not minimal to start with,
  contradiction. Then we extend the execution $\G$ by appending to
  each node labeled with a configuration $q(d_1, \ldots, d_{\#(q)})$
  the cube $\cube{\overline{\K}}$. By repeating this step for all
  leaves of a tree in $\G$, we obtain an execution of $\A$ over $w$.
  
  ``$\supseteq$'' Let $\T$ be an execution of $\A$ over $w$. We show
  that $\I_{\T}$ is a minimal interpretation such that $\I_{\T},
  \data{w} \models \pathform{\event{w}}$, by induction on $n \geq
  0$. For $n=0$, $\T$ is a cube from $\cube{\minsem{\iota}}$, by
  definition. Then $\I_{\T} \models \stamp{\iota}{0}$ and moreover, it
  is a minimal such interpretation. For the inductive case $n > 0$,
  let $w=u\cdot(a_n,\nu_n)$ for a word $u$. Let $\G$ be the
  restriction of $\T$ to $u$. Consequently, $\I_\G$ is the restriction
  of $\I_\T$ to $\stamp{Q}{\leq n-1}$. By the inductive hypothesis,
  $\I_{\G}$ is a minimal interpretation such that \(\I_{\G}, u_{\Data}
  \models \pathform{u_\Sigma}\). Since $\I_{\T}(\stamp{q}{n}) =
  \set{\tuple{d_1,\ldots,d_{\#(q)}} \mid q(d_1,\ldots,d_{\#(q)})
    \text{ labels a node on the $n$-th level in $\T$}}$, we have $\I_{\T},
  \data{w} \models \varphi$, for each subformula $\varphi = \forall y_1
  \ldots \forall y_{\#(q)} ~.~ \stamp{q}{n-1}(\vec{y})
  \rightarrow \stamp{\psi}{n}$ of $\pathform{\event{w}}$, by the
  execution semantics of $\A$. This is the case because the children
  of each node labeled with $q(d_1,\ldots,d_{\#(q)})$ on the
  $(n-1)$-th level of $\T$ form a cube from
  $\cube{\minsem{\psi}_\nu}$, where $\nu$ is a valuation that assigns
  each $y_i$ the value $d_i$ and behaves like $\data{w}$,
  otherwise. Now supppose, for a contradiction, that $\I_{\T}$ is not
  minimal and let $\J \subsetneq \I_{\T}$ be an interpretation such
  that $\J,\data{w} \models \pathform{\event{w}}$. First, we show that
  the restriction $\J'$ of $\J$ to $\bigcup_{i=0}^{n-1} \stamp{Q}{i}$
  must coincide with $\I_{\G}$. Assuming this is not the case,
  i.e.\ $\J' \subsetneq \I_{\G}$, contradicts the minimality of
  $\I_{\G}$. Then the only possibility is that $\J(\stamp{q}{n})
  \subsetneq \I_{\T}(\stamp{q}{n})$, for some $q \in Q$. Let
  $p_1(y_1,\ldots,y_{\#(p_1)}) \arrow{a_n}{} \psi_1, \ldots,
  p_k(y_1,\ldots,y_{\#(p_k)}) \arrow{a_n}{} \psi_k$ be the set of
  transition rules in which the predicate symbol $q$ occurs on the
  right-hand side. Then it must be the case that, for some node on the
  $(n-1)$-th level of $\G$, labeled with a configuration
  $p_i(d_1,\ldots,d_{\#(p_i)})$, the set of children does not form a
  minimal cube from $\cube{\minsem{\stamp{\psi_i}{n}}}$, which
  contradicts the execution semantics of $\A$. \qed}

Next, we give a logical characterization of acceptance, relative to a
given sequence of input events $\alpha \in \Sigma^*$. To this end, we
constrain the path formula $\pathform{\alpha}$ by requiring that only
final states of $\A$ occur on the last level of the execution.  The
result is the \emph{acceptance formula} for $\alpha$:
\begin{equation}\label{eq:accform}
\accform{\alpha} \isdef \pathform{\alpha} \wedge \bigwedge_{q
  \in Q \setminus F} \forall y_1 \ldots \forall y_{\#(q)} ~.~
\stamp{q}{n}(\vec{y}) \rightarrow \bot
\end{equation}
The top-level universal quantifiers from a subformula $\forall y_1
\ldots \forall y_{\#(q)} ~.~ \stamp{q}{i}(\vec{y}) \rightarrow \psi$
of $\accform{\alpha}$ will be referred to as \emph{path quantifiers},
in the following. Notice that path quantifiers are distinct from the
transition quantifiers that occur within a formula $\psi$ of a
transition rule $q(y_1,\ldots,y_{\#(q)}) \arrow{a(X)}{} \psi$ of $\A$.

The acceptance formula $\accform{\A}$ is false in every interpretation
of the predicates that assigns a non-empty set to a non-final
predicate occurring on the last level in the execution forest. The
relation between the words accepted by $\A$ and the acceptance formula
above, is formally captured by the following lemma:

\begin{lemma}\label{lemma:acceptance}
  Given an automaton $\A = \tuple{\Sigma,X,Q,\iota,F,\Delta}$, for
  every word $w \in \Sigma[X]^*$, the following are
  equivalent: \begin{compactenum}
  \item\label{it1:lemma:acceptance} there exists an interpretation
    $\I$ such that $\I,\data{w} \models \accform{\event{w}}$, 
  \item\label{it2:lemma:acceptance} $w \in \lang{\A}$. 
  \end{compactenum}
\end{lemma}
\proof{ ``$(\ref{it1:lemma:acceptance}) \Rightarrow
  (\ref{it2:lemma:acceptance})$'' Let $\I$ be an interpretation such
  that $\I,\data{w} \models \accform{\event{w}}$. By Lemma
  \ref{lemma:path-formula}, $\A$ has an execution $\T$ over $w$ such
  that $\I = \I_{\T}$. To prove that $\T$ is accepting, we show
  that \begin{inparaenum}[(i)]
    \item\label{it1:acceptance} all paths in $\T$ have length $n$ and
      that
    \item\label{it2:acceptance} the frontier of $\T$ is labeled with
      final configurations only. \end{inparaenum} First, assume that
  (\ref{it1:acceptance}) there exists a path in $\T$ of length $0 \leq
  m < n$. Then there exists a node on the $m$-th level, labeled with
  some configuration $q(d_1,\ldots,q_{\#(q)})$, that has no
  children. By the definition of the execution semantics of $\A$, we
  have $\cube{\minsem{\psi}_{\eta}} = \emptyset$, where
  $q(\vec{y}) \arrow{a_{m+1}(X)}{} \psi$ is the transition
  rule of $\A$ that applies for $q$ and $a_{m+1}$ and $\eta=
  \data{w}[y_1 \leftarrow d_1,\ldots,y_{\#(q)} \leftarrow
    d_{\#(q)}]$. Hence $\sem{\psi}_{\eta} = \emptyset$, and because
  $\I,\data{w} \models \accform{\alpha}$, we obtain that $\I,\eta \models
  q(\vec{y}) \rightarrow \stamp{\psi}{m+1}$, thus
  $\tuple{d_1,\ldots,d_{\#(q)}} \not\in \I(q)$. However, this
  contradicts the fact that $\I=\I_{\T}$ and that
  $q(d_1,\ldots,d_{\#(q)})$ labels a node of $\T$. Second, assume that
  (\ref{it2:acceptance}), there exists a frontier node of $\T$ labeled
  with a configuration $q(d_1,\ldots,d_{\#(q)})$ such that $q \in Q
  \setminus F$. Since $\I,\data{w} \models \forall y_1 \ldots \forall
  y_{\#(q)} ~.~ q(\vec{y}) \rightarrow \bot$, by a
  similar reasoning as in the above case, we obtain that
  $\tuple{d_1,\ldots,d_{\#(q)}} \not\in \I(q)$, contradiction.  

  ``$(\ref{it2:lemma:acceptance}) \Rightarrow
  (\ref{it1:lemma:acceptance})$'' Let $\T$ be an accepting execution
  of $\A$ over $w$. We prove that $\I_\T,\data{w} \models
  \accform{\event{w}}$. By Lemma \ref{lemma:path-formula}, we obtain
  $\I_\T,\data{w} \models \pathform{\event{w}}$. Since every path in
  $\T$ is of length $n$ and all nodes on the $n$-th level of $\T$ are
  labeled by final configurations, we obtain that $\I_\T,\data{w}
  \models \bigwedge_{q \in Q \setminus F} \forall y_1 \ldots \forall
  y_{\#(q)} ~.~ \stamp{q}{n}(\vec{y}) \rightarrow \bot$,
  trivially.  \qed}

As an immediate consequence, one can decide whether $\A$ accepts some
word $w$ with a given input sequence $\event{w}=\alpha$, by checking
whether $\accform{\alpha}$ is satisfiable. However, unlike
non-alternating infinite-state models of computation, such as counter
automata (nondeterministic programs with integer variables), the
satisfiability query for an acceptance (path) formula falls outside of
known decidable theories, supported by standard SMT solvers. There are
basically two reasons for this, namely\begin{inparaenum}[(i)]
\item the presence of predicate symbols, and
\item the non-trivial alternation of quantifiers.
\end{inparaenum}
To understand this point, consider for example, the decidable theory
of Presburger arithmetic \cite{Presburger29}. Adding even only one
monadic predicate symbol to it yields undecidability in the presence
of non-trivial quantifier alternation \cite{Halpern91}. However the
quantifier-free fragment of Presburger arithmetic extended with
predicate symbols can be shown to be decidable, using a Nelson-Oppen
style congruence closure argument \cite{NelsonOppen80}.

To tackle this problem, we start from the observation that acceptance
formulae have a particular form, which allows the elimination of path
quantifiers and of predicates, by a couple of
satisfiability-preserving transformations. The result of applying
these transformations is a formula with no predicate symbols, whose
only quantifiers are those introduced by the transition rules of the
automaton, referred to as transition quantifiers. We shall further
assume (\S\ref{sec:emptiness}) that the first order theory of the data
sort $\Data$ has quantifier elimination, which allows to effectively
decide the satisfiability of such formulae.

For the time being, let us formally define the elimination of
transition quantifiers and predicates, respectively. Consider a given
sequence of input events $\alpha = a_1 \ldots a_n$ and denote by
$\alpha_i$ the prefix $a_1 \ldots a_i$ of $\alpha$, for $i \in [n]$,
where $\alpha_0=\epsilon$.

\begin{definition}\label{def:quantform}
  Let $\quantformn{\alpha_0}, \ldots, \quantformn{\alpha_n}$ be the
  sequence of formulae defined by $\quantformn{\alpha_0} \isdef
  \stamp{\iota}{0}$ and, for all $i \in [1,n]$:
  \vspace*{-.5\baselineskip}
  \[\quantformn{\alpha_i} \isdef \quantformn{\alpha_{i-1}} \wedge
  \bigwedge_{\begin{array}{l}
      \scriptstyle{\stamp{q}{i-1}(t_1,\ldots,t_{\#(q)}) \text{ occurs in } \quantformn{\alpha_{i-1}}} \\
      \scriptstyle{q(y_1,\ldots,y_{\#(q)}) \arrow{a_i(X)}{} \psi \in \Delta} 
  \end{array}} \stamp{q}{i-1}(t_1,\ldots,t_{\#(q)}) \rightarrow 
    \stamp{\psi}{i}[t_1/y_1, \ldots, t_{\#(q)}/y_{\#(q)}]
      \]  
      \vspace*{-.5\baselineskip}
      We write
    $\quantform{\alpha}$ for the prenex normal form of the formula: 
      \[\quantformn{\alpha_n} \wedge \bigwedge_{\begin{array}{l}
          \scriptstyle{\stamp{q}{n}(t_1,\ldots,t_{\#(q)}) \text{ occurs in } \quantformn{\alpha_n}} \\
          \scriptstyle{q \in Q \setminus F}
      \end{array}} \stamp{q}{n}(t_1,\ldots,t_{\#(q)}) \rightarrow \bot \enspace.\]
\end{definition}
Observe that $\quantform{\alpha}$ contains no path quantifiers, as
required. On the other hand, the scope of the transition quantifiers
in $\quantform{\alpha}$ exceeds the right-hand side formulae from the
transition rules, as shown by the following example. 

\begin{example}\label{ex:quant-elim}
  Consider the automaton $\A = \tuple{\set{a_1,a_2}, \set{x},
    \set{q,q_f}, \iota, \set{q_f}, \Delta}$, where:
  \[\begin{array}{rcl}
  \iota & = & \exists z ~.~ z \geq 0 \wedge q(z) \\
  \Delta & = & \set{
    q(y) \arrow{a_1(x)}{} x\geq 0 \wedge \forall z ~.~ z \leq y \rightarrow q(x+z),~
    q(y) \arrow{a_2(x)}{} y<0 \wedge q_f(x+y)
  }
  \end{array}\]
  For the input event sequence $\alpha = a_1a_2$, the acceptance
  formula is:
  \[\begin{array}{rcl}
  \accform{\alpha} & = & \exists z ~.~ z \geq 0 \wedge \stamp{q}{0}(z) ~\wedge \\
  && \forall y ~.~ \stamp{q}{0}(y) \rightarrow [\stamp{x}{1}\geq0 \wedge \forall z ~.~ z\geq y 
    \rightarrow \stamp{q}{1}(\stamp{x}{1}+z)] ~\wedge \\
  && \forall y ~.~ \stamp{q}{1}(y) \rightarrow [y<0 \wedge \stamp{q_f}{2}(\stamp{x}{2}+y)]
  \end{array}\]
  The result of eliminating the path quantifiers, in prenex normal form, is shown below:
  \[\begin{array}{rcl}
  \quantform{\alpha} & = & \exists z_1\forall z_2 ~.~ z_1 \geq 0 \wedge \stamp{q}{0}(z_1) ~\wedge \\
  && [\stamp{q}{0}(z_1) \rightarrow \stamp{x}{1} \geq 0 \wedge (z_2 \geq z_1 \rightarrow \stamp{q}{1}(\stamp{x}{1}+z_2))] ~\wedge \\
  && [\stamp{q}{1}(\stamp{x}{1}+z_2) \rightarrow \stamp{x}{1}+z_2 < 0 \wedge \stamp{q_f}{2}(\stamp{x}{2}+\stamp{x}{1}+z_2)] 
  \enspace\enspace\hfill\blacksquare
  \end{array}\]
\end{example}

The next lemma establishes a formal relation between the
satisfiability of an acceptance formula $\accform{\alpha}$ and that of
the formula $\quantform{\alpha}$, obtained by eliminating the path
quantifiers from $\accform{\alpha}$.

\begin{lemma}\label{lemma:quant}
  For any input event sequence $\alpha=a_1\ldots a_n$ and each
  valuation $\nu : \stamp{X}{\leq n} \rightarrow \Data$, the following
  hold: \begin{compactenum}
  \item\label{it1:quant} for all interpretations $\I$, if $\I,\nu
    \models \accform{\alpha}$ then $\I,\nu \models
    \quantform{\alpha}$. 
  \item\label{it2:quant} if there exists an interpretation $\I$ such
    that $\I,\nu \models \quantform{\alpha}$ then there exists an
    interpretation $\J \subseteq \I$ such that $\J,\nu \models
    \accform{\alpha}$. 
  \end{compactenum}
\end{lemma}
\proof{(\ref{it1:quant}) Trivial, since every subformula
  $q(t_1,\ldots,t_{\#(q)}) \rightarrow
  \psi[t_1/y_1,\ldots,t_{\#(q)}/y_{\#(q)}]$ of $\quantform{\alpha}$ is
  entailed by a subformula $\forall y_1 \ldots \forall y_{\#(q)} ~.~
  q(y_1, \ldots, y_{\#(q)}) \rightarrow \psi$ of $\accform{\alpha}$.

  \noindent(\ref{it2:quant}) By repeated applications of the following fact: 

  \begin{fact}
    Given formulae $\phi$ and $\psi$, such that no predicate atom with
    predicate symbol $q$ occurs in $\psi(y_1,\ldots,y_{\#(q)})$, for
    each valuation $\nu$, if there exists an interpretation $\I$ such
    that $\I,\nu \models \phi \wedge
    \bigwedge_{q(t_1,\ldots,t_{\#(q)}) \text{ occurs in } \phi}
    q(t_1,\ldots,t_{\#(q)}) \rightarrow
    \psi[t_1/y_1,\ldots,t_{\#(q)}/y_{\#(q)}]$ then there exists a
    valuation $\J$ such that $\J(q) \subseteq \I(q)$ and $\J(q') =
    \I(q')$ for all $q' \in Q \setminus \set{q}$ and $\J,\nu \models
    \phi \wedge \forall y_1 \ldots \forall y_{\#(q)} ~.~
    q(y_1,\ldots,y_{\#(q)}) \rightarrow \psi$.
  \end{fact}
  \proof{Assume w.l.o.g. that $\phi$ is quantifier free. The proof can
    be easily generalized to the case $\phi$ has quantifiers. Let
    $\J(q) = \set{\langle t_1^\nu, \ldots, t_{\#(q)}^\nu \rangle \in
      \I(q) \mid q(t_1,\ldots,t_{\#(q)}) \text{ occurs in } \phi}$ and
    $\J(q') = \I(q')$ for all $q' \in Q \setminus \set{q}$. Since
    $\I,\nu \models \phi$, we obtain that also $\J,\nu \models \phi$
    because the tuples of values in $\I(q) \setminus \J(q)$ are not
    interpretations of terms that occur within subformulae
    $q(t_1,\ldots,t_{\#(q)})$ of $\phi$. Moreover,
    $\bigwedge_{q(t_1,\ldots,t_{\#(q)}) \text{ occurs in } \phi}
    q(t_1,\ldots,t_{\#(q)}) \rightarrow
    \psi[t_1/y_1,\ldots,t_{\#(q)}/y_{\#(q)}]$ and $\forall y_1 \ldots
    \forall y_{\#(q)} ~.~ q(y_1,\ldots,y_{\#(q)}) \rightarrow \psi$
    are equivalent under $\J$, thus $\J,\nu \models \forall y_{\#(q)}
    ~.~ q(y_1,\ldots,y_{\#(q)}) \rightarrow \psi$, as required. \qed}
  
  \noindent This concludes the proof. \qed}

We proceed with the elimination of predicate atoms from
$\quantform{\alpha}$, defined below. 

\begin{definition}\label{def:substform}
  Let $\substformn{\alpha_0}, \ldots, \substformn{\alpha_n}$ be the
  sequence of formulae defined by $\substformn{\alpha_0} \isdef
  \stamp{\iota}{0}$ and, for all $i \in [1,n]$,
  $\substformn{\alpha_i}$ is obtained by replacing each occurrence of
  a predicate atom $\stamp{q}{i-1}(t_1,\ldots,t_{\#(q)})$ in
  $\substformn{\alpha_{i-1}}$ with the formula
  $\stamp{\psi}{i}[t_1/y_1,\ldots,t_{\#(q)}/y_{\#(q)}]$, where
  $q(\vec{y}) \arrow{a_{i}(X)}{} \psi \in \Delta$. We write
  $\substform{\alpha}$ for the formula obtained by replacing, in
  $\substformn{\alpha}$, each occurrence of a predicate
  $\stamp{q}{n}$, such that $q \in Q \setminus F$ (resp. $q \in F$),
  by $\bot$ (resp. $\top$).
\end{definition}

\begin{example}[Contd. from Example \ref{ex:quant-elim}]\label{ex:pred-elim}
  The result of the elimination of predicate atoms from the acceptance
  formula in Example \ref{ex:quant-elim} is shown below:
  \[\substform{\alpha} = 
  \exists z_1 \forall z_2 ~.~ z_1 \geq 0 \wedge 
          [\stamp{x}{1} \geq 0 \wedge (z_2 \geq z_1 \rightarrow
            \stamp{x}{1}+z_2 < 0)]\] Since this formula is
          unsatisfiable, by Lemma \ref{lemma:quant-pred-acceptance}
          below, no word $w$ with input event sequence $\event{w} =
          a_1a_2$ is accepted by the automaton $\A$ from Example
          \ref{ex:quant-elim}. \hfill$\blacksquare$
\end{example}

At this point, we prove the formal relation between the satisfiability
of the formulae $\quantform{\alpha}$ and $\substform{\alpha}$. Since
there are no occurrences of predicates in $\substform{\alpha}$, for
each valuation $\nu : \stamp{X}{\leq n} \rightarrow \Data$, there
exists an interpretation $\I$ such that $\I,\nu \models
\substform{\alpha}$ if and only if $\J,\nu \models
\substform{\alpha}$, for every interpretation $\J$. In this case we
omit $\I$ and simply write $\nu \models \substform{\alpha}$.

\begin{lemma}\label{lemma:subst}
  For any input event sequence $\alpha=a_1\ldots a_n$ and each
  valuation $\nu : \stamp{X}{\leq n} \rightarrow \Data$, there exists
  a valuation $\I$ such that $\I,\nu \models \quantform{\alpha}$ if
  and only if $\nu \models \substform{\alpha}$. 
\end{lemma}
\proof{ By induction on $n\geq0$. The base case $n=0$ is trivial,
  since $\quantform{A} = \substform{A} = \stamp{\iota}{0}$. For the
  induction step, we rely on the following fact: 

  \begin{fact}\label{fact:subst}
    Given formulae $\phi$ and $\psi$, such that $\phi$ is positive
    $q(t_1,\ldots,t_{\#(q)})$ is the only one occurrence of the
    predicate symbol $q$ in $\phi$ and no predicate atom with
    predicate symbol $q$ occurs in $\psi(y_1,\ldots,y_{\#(q)})$, for
    each interpretation $\I$ and each valuation $\nu$, we have:
    \[\begin{array}{l} 
    \I,\nu \models \phi \wedge q(t_1,\ldots,t_{\#(q)}) \rightarrow
    \psi[t_1/y_1,\ldots,t_{\#(q)}/y_{\#(q)}] \iff \\ \nu \models
    \phi[\psi[t_1/y_1,\ldots,t_{\#(q)}/y_{\#(q)}]/q(t_1,\ldots,t_{\#(q)})]\enspace.
    \end{array}\]
  \end{fact}
  \proof{ We assume w.l.o.g. that $\phi$ is quantifier-free. The proof
    can be easily generalized to the case $\phi$ has quantifiers.

    \noindent''$\Rightarrow$'' 
    We distinguish two cases: \begin{compactitem}
    \item if $\tuple{t_1^\nu, \ldots, t_{\#(q)}^\nu} \in \I(q)$ then
      $\I,\nu \models \psi[t_1/y_1,\ldots,t_{\#(q)}/y_{\#(q)}]$. Since
      $\phi$ is positive, replacing $q(t_1,\ldots,t_{\#(q)})$ with
      $\psi[t_1/y_1,\ldots,t_{\#(q)}/y_{\#(q)}]$ does not change the
      truth value of $\phi$ under $\nu$, thus $\nu \models
      \phi[\psi[t_1/y_1,\ldots,t_{\#(q)}/y_{\#(q)}]/q(t_1,\ldots,t_{\#(q)})]$.
    \item else, $\tuple{t_1^\nu, \ldots, t_{\#(q)}^\nu} \not\in
      \I(q)$, thus $\nu \models
      \phi[\bot/q(t_1,\ldots,t_{\#(q)})]$. Since $\phi$ is positive
      and $\bot$ entails $\psi[t_1/y_1,\ldots,t_{\#(q)}/y_{\#(q)}]$,
      we obtain $\nu \models
      \phi[\psi[t_1/y_1,\ldots,t_{\#(q)}/y_{\#(q)}]/q(t_1,\ldots,t_{\#(q)})]$
      by monotonicity. 
    \end{compactitem}

    \noindent``$\Leftarrow$'' Let $\I$ be any interpretation such that
    $\I(q) = \set{\tuple{t_1^\nu, \ldots, t_{\#(q)}^\nu} \mid \nu
      \models \psi[t_1/y_1,\ldots,t_{\#(q)}/y_{\#(q)}]}$. We
    distinguish two cases: \begin{compactitem}
      \item if $\I(q) \not \emptyset$ then $\I,\nu \models
        q(t_1,\ldots,t_{\#(q)})$ and $\nu \models \psi[t_1/y_1,
          \ldots, t_{\#(q)}/y_{\#(q)}]$. Thus replacing $\psi[t_1/y_1,
          \ldots, t_{\#(q)}/y_{\#(q)}]$ by $q(t_1,\ldots,t_{\#(q)})$
        does not change the truth value of $\phi$ under $\I$ and
        $\nu$, and we obtain $\I,\nu \models \phi$. Moreover, $\I,\nu
        \models \psi[t_1/y_1, \ldots, t_{\#(q)}/y_{\#(q)}]$ implies
        $\I,\nu \models q(t_1,\ldots,t_{\#(q)}) \rightarrow
        \psi[t_1/y_1, \ldots, t_{\#(q)}/y_{\#(q)}]$.
      \item else $\I(q) = \emptyset$, hence $\nu \not\models
        \psi[t_1/y_1, \ldots, t_{\#(q)}/y_{\#(q)}]$, thus $\nu \models
        \phi[\bot/q(t_1,\ldots,t_{\#(q)})]$. Because $\phi$ is
        positive, we obtain $\I,\nu \models \phi$ by monotonicity. But
        $\I,\nu \models q(t_1,\ldots,t_{\#(q)}) \rightarrow
        \psi[t_1/y_1, \ldots, t_{\#(q)}/y_{\#(q)}]$ trivially, because
        $\I,\nu \not\models q(t_1,\ldots,t_{\#(q)})$.\qed
    \end{compactitem}}
This concludes the proof. \qed}

Finally, we define the acceptance of a word with a given input event
sequence by means of a formula in which no predicate atom occurs. As
previously discussed, several decidable theories, such as Presburger
arithmetic, become undecidable if predicate atoms are added to
them. Therefore, the result below makes a step forward towards
deciding whether the automaton accepts a word with a given input
sequence, by reducing this problem to the satisfiability of a
quantified formula without predicates.

\begin{lemma}\label{lemma:quant-pred-acceptance}
  Given an automaton $\A = \tuple{\Sigma,X,Q,\iota,F,\Delta}$, for
  every word $w \in \Sigma[X]^*$, we have $\data{w} \models
  \substform{\event{w}}$ if and only if $w \in \lang{\A}$.
\end{lemma}
\proof{ By Lemma \ref{lemma:acceptance}, $w \in \lang{\A}$ if and only
  if $\I,\data{w} \models \accform{\event{w}}$, for some
  interpretation $\I$. By Lemma \ref{lemma:quant}, there exists an
  interpretation $\I$ such that $\I,\data{w} \models
  \accform{\event{w}}$ if and only if there exists an interpretation
  $\J$ such that $\J,\nu \models \quantform{\event{w}}$. By Lemma
  \ref{lemma:subst}, there exists an interpretation $\J$ such that
  $\J,\nu \models \quantform{\event{w}}$ if and only if $\nu \models
  \substform{\event{w}}$. \qed}

\subsection{Closure Properties}
\label{sec:closure}

Given a positive formula $\phi$, we define the \emph{dual} formula
$\dual{\phi}$ recursively as follows:
\[\begin{array}{rclcrclcrcl}
\dual{(\phi_1 \vee \phi_2)} & = & \dual{\phi_1} \wedge \dual{\phi_2} && 
\dual{(\phi_1 \wedge \phi_2)} & = & \dual{\phi_1} \vee \dual{\phi_2} &&
\dual{(t \teq s)} & = & \neg(t \teq s) \\
\dual{(\exists x ~.~ \phi_1)} & = & \forall x ~.~ \dual{\phi_1} && 
\dual{(\forall x ~.~ \phi_1)} & = & \exists x ~.~ \dual{\phi_1} && 
\dual{(\neg(t \teq s))} & = & t \teq s \\
\dual{(q(x_1,\ldots,x_{\#(q)}))} & = & q(x_1,\ldots,x_{\#(q)})
\end{array}\]
Observe that, because predicate atoms do not occur negated in $\phi$,
there is no need to define dualization for formulae of the form $\neg
q(x_1,\ldots,x_{\#(q)})$. The following theorem shows closure of
automata under all boolean operations:


\begin{theorem}\label{thm:closure}
  Given automata $\A_i = \tuple{\Sigma,X,Q_i,\iota_i,F_i,\Delta_i}$,
  for $i=1,2$, such that $Q_1 \cap Q_2 = \emptyset$, the following hold: 
  \begin{compactenum}
    \item\label{it1:thm:closure} $\lang{\A_\cap} = \lang{\A_1} \cap \lang{\A_2}$, where
      $\A_\cap = \tuple{\Sigma,X,Q_1 \cup Q_2, \iota_1 \wedge \iota_2,
      F_1 \cup F_2, \Delta_1 \cup \Delta_2}$, 
    \item\label{it2:thm:closure} $\lang{\overline{\A_i}} = \Sigma[X]^*
      \setminus \lang{\A_i}$, where $\overline{\A_i} =
      \tuple{\Sigma,X,Q_i,\dual{\iota},Q_i\setminus
        F_i,\dual{\Delta}_i}$ and, for $i=1,2$: \[\dual{\Delta}_i =
      \set{q(\vec{y}) \arrow{a(X)}{} \dual{\psi} \mid
        q(\vec{y}) \arrow{a(X)}{} \psi \in
        \Delta_i}\enspace.\]
  \end{compactenum}
  Moreover, $\len{\A_\cap} = \bigO{\len{\A_1}+\len{\A_2}}$ and
  $\len{\overline{\A_i}} = \bigO{\len{\A_i}}$, for all $i=1,2$.
\end{theorem}
\proof{ (\ref{it1:thm:closure}) ``$\subseteq$'' Let $w \in
  \lang{\A_\cap}$ be a word and $\T$ be an execution of $\A_\cap$ over
  $w$. Since $Q_1 \cap Q_2 = \emptyset$, it is possible to partition
  $\T$ into $\T_1$ and $\T_2$ such that the roots of $\T_i$ form a
  cube from $\cube{\minsem{\iota_i}}$, for all $i=1,2$. Because
  $\Delta_1 \cap \Delta_2 = \emptyset$, by induction on
  $\len{w}\geq0$, one shows that $\T_i$ is an execution of $\A_i$ over
  $w$, for all $i=1,2$. Finally, because $\T$ is accepting, we obtain
  that $\T_1$ and $\T_2$ are accepting, respectively, hence $w \in
  \lang{\A_1} \cap \lang{\A_2}$. ``$\supseteq$'' Let $w \in
  \lang{\A_1} \cap \lang{\A_2}$ and let $\T_i$ an accepting execution
  of $\A_i$ over $w$, for all $i=1,2$. We show that $\T_1 \cup \T_2$
  is an execution of $\A_\cap$ over $w$, by induction on
  $\len{w}\geq0$. For the base case $\len{w}=0$, we have $\T_i \in
  \cube{\minsem{\iota_i}}$ for all $i=1,2$ and since $Q_1 \cap Q_2 =
  \emptyset$, we have $\T_1 \cup \T_2 \in \cube{\minsem{\iota_1 \wedge
      \iota_2}}$. The induction step follows as a consequence of the
  fact that $\Delta_1 \cup \Delta_2$ is the set of transition rules of
  $\A_\cap$. Finally, since both $\T_1$ and $\T_2$ are accepting,
  $\T_1 \cup \T_2$ is accepting as well. Moreover, we have: 
  \[
  \len{\A_\cap} = \len{\iota_1 \wedge \iota_2} + \sum_{q(\vec{y})
    \arrow{a(X)}{} \psi \in \Delta_1 \cup \Delta_2} \len{\psi} =
  1 + \len{\iota_1} + \len{\iota_2} + \sum_{q(\vec{y})
    \arrow{a(X)}{} \psi \in \Delta_1} \len{\psi} + \sum_{q(\vec{y})
    \arrow{a(X)}{} \psi \in \Delta_2} \len{\psi} \enspace. 
  \]

  \noindent(\ref{it2:thm:closure}) Let $w \in \Sigma[X]^*$ be a
  word. We denote by $\accformA{\A_1}{\event{w}}$ and
  $\substformA{\A_1}{\event{w}}$
  [resp. $\accformA{\overline{\A}_1}{\event{w}}$ and
    $\substformA{\overline{\A}_1}{\event{w}}$] the formulae
  $\accform{\event{w}}$ and $\substform{\event{w}}$ for $\A_1$ and
  $\overline{\A}_1$, respectively. It is enough to show that
  $\substformA{\overline{\A}_1}{\event{w}} =
  \neg\substformA{\A_1}{\event{w}}$ and apply Lemma
  \ref{lemma:quant-pred-acceptance} to prove that $w \in \lang{\A_1}
  \iff w \not\in \lang{\overline{\A}_1}$. Since the choice of $w$ was
  arbitrary, this proves $\lang{\overline{\A}_1} = \Sigma[X]^*
  \setminus \lang{\A_1}$. By induction on the number of predicate
  atoms in $\accformA{\A_1}{\event{w}}$ that are replaced during the
  generation of $\substformA{\A_1}{\event{w}}$. The proof relies on
  the following fact:

  \begin{fact}\label{fact:dual-neg}
    Let $\phi$ be a positive formula and let $q(t_1,\ldots,t_{\#(q)})$
    be the only occurrence of a predicate symbol within $\phi$. Then,
    every formula $\phi$ with no predicate occurrences:
    $\neg\phi[\psi[t_1/y_1,\ldots,t_{\#(q)}/y_{\#(q)}]/q(t_1,\ldots,t_{\#(q)})]
    \equiv
    \dual{\phi}[\neg\psi[t_1/y_1,\ldots,t_{\#(q)}/y_{\#(q)}]/q(t_1,\ldots,t_{\#(q)})]$.
  \end{fact}
  \proof{By induction on the structure of $\phi$. \qed}}

\section{The Emptiness Problem}
\label{sec:emptiness}

The problem of checking emptiness of a given automaton is undecidable,
even for automata with predicates of arity two, whose transition rules
use only equalities and disequalities, having no transition
quantifiers \cite{Farzan15}. Since even such simple classes of
alternating automata have no general decision procedure for emptiness,
we use an abstraction-refinement semi-algorithm based on \emph{lazy
  annotation} \cite{McMillan06,McMillan14}.

In a nutshell, a lazy annotation procedure systematically explores the
set of execution paths (in our case, sequences of input events) in
search of an accepting execution. Each path has a corresponding path
formula that defines all words accepted along that path. If the path
formula is satisfiable, the automaton accepts a word. Otherwise, the
path is said to be \emph{spurious}. When a spurious path is
encountered, the search backtracks and the path is annotated with a
set of learned facts, that marks this path as infeasible. The
semi-algorithm uses moreover a coverage relation between paths,
ensuring that the continuations of already covered paths are never
explored. Sometimes this coverage relation provides a sound
termination argument, when the automaton is empty.

We check emptiness of first order alternating automata using a version
of the \impact~ lazy annotation semi-algorithm \cite{McMillan06}. An
analogous procedure is given in \cite{IosifXu18}, for a simpler model
of alternating automata, that uses only predicates or arity zero
(booleans) and no transition quantifiers. For simplicity, we do not
present the details of this algorithm and shall content ourselves of
several high-level definitions.

Given a finite input event alphabet $\Sigma$, for two sequences
$\alpha, \beta \in \Sigma^*$, we say that $\alpha$ is a prefix of
$\beta$, written $\alpha \prefix \beta$, if $\alpha=\beta\gamma$ for
some sequence $\gamma\in\Sigma^*$. A set $S$ of sequences
is: \begin{compactitem}
\item \emph{prefix-closed} if for each $\alpha \in S$, if $\beta \prefix
\alpha$ then $\beta \in S$, and
\item \emph{complete} if for each $\alpha \in S$, there exists $a \in
  \Sigma$ such that $\alpha a \in S$ if and only if $\alpha b \in S$
  for all $b \in \Sigma$. 
\end{compactitem}
Observe that a prefix-closed set is the backbone of a tree whose edges
are labeled with input events. If the set is, moreover, complete, then
every node of the tree has either zero successors, in which case it is
called a \emph{leaf}, or it has a successor edge labeled with $a$ for
each input event $a \in \Sigma$. 

\begin{definition}\label{def:unfolding}
  An \emph{unfolding} of an automaton $\A =
  \tuple{\Sigma,X,Q,\iota,F,\Delta}$ is a finite partial mapping $U :
  \Sigma^* \rightharpoonup_{\mathit{fin}} \posforms(Q,\emptyset)$,
  such that: \begin{compactenum}
  \item\label{it1:unfolding} $\dom(U)$ is a finite prefix-closed
    complete set,
  \item\label{it2:unfolding} $U(\epsilon) = \iota$, and
  \item\label{it3:unfolding} for each sequence $\alpha a \in \dom(U)$,
    such that $\alpha \in \Sigma^*$ and $a \in \Sigma$:
    \[\stamp{U(\alpha)}{0} \wedge
    \bigwedge_{q(\vec{y}) \arrow{a(X)}{} \psi} \forall y_1 \ldots
    \forall y_{\#q} ~.~ \stamp{q}{0}(\vec{y}) \rightarrow
    \stamp{\psi}{1} \models \stamp{U(\alpha a)}{1}\]
  \end{compactenum}
  Moreover, $U$ is \emph{safe} if for each $\alpha \in \dom(U)$, the
  formula $U(\alpha) \wedge \bigwedge_{q \in Q \setminus F} \forall
  y_1 \ldots \forall_{y_{\#(q)}} ~.~ q(\vec{y}) \rightarrow \bot$ is
  unsatisfiable. 
\end{definition}

Lazy annotation semi-algorithms \cite{McMillan06,McMillan14} build
unfoldings of automata trying to discover counterexamples for
emptiness. If the automaton $\A$ in question is non-empty, a
systematic enumeration of the input event sequences\footnote{For
  instance, using breadth-first search.} from $\Sigma^*$ will suffice
to discover a word $w \in \lang{\A}$, provided that the first order
theory of the data domain $\Data$ is decidable (Lemma
\ref{lemma:acceptance}). However, if $\lang{\A} = \emptyset$, the
enumeration of input event sequences may, in principle, run
forever. The typical way of fighting this divergence problem is to
define a \emph{coverage} relation between the nodes of the unfolding
tree.

\begin{definition}\label{def:coverage}
  Given an unfolding $U$ of an automaton $\A =
  \tuple{\Sigma,X,Q,\iota,F,\Delta}$ a node $\alpha \in \dom(U)$ is
  \emph{covered} by another node $\beta \in \dom(U)$, denoted $\alpha
  \cover \beta$, if and only if there exists a node $\alpha' \prefix
  \alpha$ such that $U(\alpha') \models U(\beta)$. Moreover, $U$ is
  \emph{closed} if and only if every leaf from $\dom(U)$ is covered by
  an uncovered node.
\end{definition}


A lazy annotation semi-algorithm will stop and report emptiness
provided that it succeeds in building a closed and safe unfolding of
the automaton. Notice that, by Definition \ref{def:coverage}, for any
three nodes of an unfolding $U$, say $\alpha,\beta,\gamma \in
\dom(U)$, if $\alpha \prec \beta$ and $\alpha \cover \gamma$, then
$\beta \cover \gamma$ as well. As we show next (Theorem
\ref{thm:soundness}), there is no need to expand covered nodes,
because, intuitively, there exists a word $w \in \lang{\A}$ such that
$\alpha \preceq \event{w}$ and $\alpha \cover \gamma$ only if there
exists another word $u \in \lang{\A}$ such that $\gamma \preceq
\event{u}$. Hence, exploring only those input event sequences that are
continuations of $\gamma$ (and ignoring those of $\alpha$) suffices in
order to find a counterexample for emptiness, if one exists.

An unfolding node $\alpha \in \dom(U)$ is said to be \emph{spurious}
if and only if $\accform{\alpha}$ is unsatisfiable. In this case, we
change (refine) the labels of (some of the) prefixes of $\alpha$ (and
that of $\alpha$), such that $U(\alpha)$ becomes $\bot$, thus
indicating that there is no real execution of the automaton along that
input event sequence. As a result of the change of labels, if a node
$\gamma \prefix \alpha$ used to cover another node from $\dom(U)$, it
might not cover it with the new label. Therefore, the coverage
relation has to be recomputed after each refinement of the
labeling. The semi-algorithm stops when (and if) a safe complete
unfolding has been found. For a detailed presentation of the emptiness
procedure, we refer to \cite{IosifXu18}.

\begin{theorem}\label{thm:soundness}
  If an automaton $\A$ has a nonempty safe closed unfolding then
  $\lang{A} = \emptyset$.
\end{theorem}
\proof{ Let $U$ be a safe and complete unfolding of $\A$, such that
  $\dom(U) \neq \emptyset$. Suppose, by contradiction, that there
  exists a word $w \in \lang{A}$ and let $\alpha \isdef
  \event{w}$. Since $w \in \lang{A}$, by Lemma \ref{lemma:acceptance},
  there exists an interpretation $\I$ such that $\I,\data{w} \models
  \accform{\alpha}$. Assume first that $\alpha \in \dom(U)$.  In this
  case, one can show, by induction on the length $n\geq0$ of $w$, that
  $\pathform{\alpha} \models \stamp{U(\alpha)}{n}$, thus $\I,\data{w}
  \models \stamp{U(\alpha)}{n}$. Since $\I,\data{w} \models
  \accform{\alpha}$, we have $\I,\data{w} \models \bigwedge_{q \in Q
    \setminus F} \forall y_1 \ldots \forall y_{\#(q)} ~.~
  \stamp{q}{n}(\vec{y}) \rightarrow \bot$, hence $\stamp{U(\alpha)}{n}
  \wedge \bigwedge_{q \in Q \setminus F} \forall y_1 \ldots \forall
  y_{\#(q)} ~.~ \stamp{q}{n}(\vec{y}) \rightarrow \bot$. By renaming
  $\stamp{q}{n}$ with $q$ in the previous formula, we obtain
  $U(\alpha) \wedge \forall y_1 \ldots \forall y_{\#(q)} ~.~
  q(\vec{y}) \rightarrow \bot$ is satisfiable, thus $U$ is not safe,
  contradiction.

  We proceed thus under the assumption that $\alpha \not\in
  \dom(U)$. Since $\dom(U)$ is a nonempty prefix-closed set, there
  exists a strict prefix $\alpha'$ of $\alpha$ that is a leaf of
  $\dom(U)$. Since $U$ is closed, the leaf $\alpha'$ must be covered
  and let $\alpha_1 \prefix \alpha' \prefix \alpha$ be a node such
  that $U(\alpha_1) \models U(\beta_1)$, for some uncovered node
  $\beta_1 \in \dom(U)$.
  Let $\gamma_1$ be the unique sequence such that $\alpha_1\gamma_1 =
  \alpha$. By Definition \ref{def:coverage}, since $\alpha_1 \cover
  \beta_1$ and $\event{w} = \alpha_1\gamma_1 \in \lang{\A}$, there
  exists a word $w_1$ and a cube $c_1 \in \cube{\sem{U(\alpha_1)}}
  \subseteq \cube{\sem{U(\beta_1)}}$, such that $\event{w_1} =
  \gamma_1$ and $\A$ accepts $w_1$ starting with $c_1$. If
  $\beta_1\gamma_1 \in \dom(U)$, we obtain a contradiction by a
  similar argument as above. Hence $\beta_1\gamma_1 \not\in \dom(U)$
  and there exists a leaf of $\dom(U)$ which is also a prefix of
  $\beta_1\gamma_1$. Since $U$ is closed, this leaf is covered by an
  uncovered node $\beta_2 \in \dom(U)$ and let $\alpha_2 \in \dom(U)$
  be the minimal (in the prefix partial order) node such that $\beta_1
  \prefix \alpha_2 \prefix \beta_1\gamma_1$ and $\alpha_2 \cover
  \beta_2$. Let $\gamma_2$ be the unique sequence such that
  $\alpha_2\gamma_2 = \beta_1\gamma_1$. Since $\beta_1$ is uncovered,
  we have $\beta_1 \neq \alpha_2$ and thus $\len{\gamma_1} >
  \len{\gamma_2}$. By repeating the above reasoning for $\alpha_2$,
  $\beta_2$ and $\gamma_2$, we obtain an infinite sequence
  $\len{\gamma_1} > \len{\gamma_2} > \ldots$, which is again a
  contradiction. \qed}

As mentioned above, we check emptiness of first order alternating
automata using the same method previously used to check emptiness of a
simpler model of alternating automata, which uses boolean constants
for control states and whose transition rules have no quantifiers
\cite{IosifXu18}. The higher complexity of the automata model
considered here, manifests itself within the interpolant generation
procedure, used to refine the labeling of the unfolding. We discuss
generation of interpolants in the next section.

\section{Interpolant Generation}\label{sec:interpolants}

Typically, when checking the unreachability of a set of program
configurations \cite{McMillan06}, the interpolants used to annotate
the unfolded control structure are assertions about the values of the
program variables in a given control state, at a certain step of an
execution. However, in an alternating model of computation, it is
useful to distinguish between
\begin{inparaenum}[(i)]
\item locality of interpolants w.r.t. a given control state (control
  locality) and
\item locality w.r.t. a given time stamp (time locality).
\end{inparaenum} 
In logical terms, \emph{control-local} interpolants are defined by
formulae involving a single predicate symbol, whereas
\emph{time-local} interpolants involve only predicates $\stamp{q}{i}$
and variables $\stamp{x}{i}$, for a single $i \geq 0$.

\paragraph{Remark}
When considering an alternating model of computation, control-local
interpolants are not always enough to prove emptiness, because of the
synchronization of several branches of the computation on the same
sequence of input values. Consider, for instance, an automaton with
the following transition rules and final state $q_f$:
\[\begin{array}{rclcrclcrcl}
q_0(y) & \arrow{a(x)}{} & q_1(y+x) \wedge q_2(y-x) & \hspace*{2mm} & 
q_1(y) & \arrow{a(x)}{} & y+x > 0 \wedge q_f & \hspace*{2mm} & 
q_1(y) & \arrow{a(x)}{} & q_1(y+x) \\ 
q_2(y) & \arrow{a(x)}{} & y-x > 0 \wedge q_f && 
q_2(y) & \arrow{a(x)}{} & q_2(y-x)
\end{array}\]
Started in an initial configuration $q_0(0)$ with an input word
$(a,\nu_1) \ldots (a,\nu_{n-1}) (a,\nu_n)$, such that $\nu_i(x)=k_i$,
the automaton executes as follows:
\[\begin{array}{c}
q_0(0) \arrow{(a,\nu_1)}{} \set{q_1(k_1),q_2(-k_1)} \ldots \arrow{(a,\nu_{n-1})}{} 
\set{q_1(\sum_{i=1}^{n-1} k_i),q_2(-\sum_{i=1}^{n-1} k_i)}
\arrow{(a,\nu_n)}{} \emptyset
\end{array}\]
An overapproximation of the set of cubes generated after one or more
steps is defined by the formula: $\exists x_1 \exists x_2 ~.~ q_1(x_1)
\wedge q_2(x_2) \wedge x_1+x_2 \teq 0$. Observe that a control-local
formula using one occurrence of a predicate would give a too rough
overapproximation of this set, unable to prove the emptiness of the
automaton. \hfill$\blacksquare$

First, let us give the formal definition of the class of interpolants
we shall work with. Given a formula $\phi$, the \emph{vocabulary} of
$\phi$, denoted $\voc{\phi}$ is the set of predicate symbols $q \in
\stamp{Q}{i}$ and variables $x \in \stamp{X}{i}$, occurring in $\phi$,
for some $i\geq0$. For a term $t$, its vocabulary $\voc{t}$ is the set
of variables that occur in $t$. Observe that quantified variables and
the interpreted function symbols of the data
theory\footnote{E.g.,\ the arithmetic operators of addition and
  multiplication, when $\Data$ is the set of integers.}  do not belong
to the vocabulary of a formula. By $\pset{+}{\phi}$ [$\pset{-}{\phi}$]
we denote the set of predicate symbols that occur in $\phi$ under an
even [odd] number of negations.

\begin{definition}[\cite{Lyndon59}]\label{def:lyndon-interpolant}
Given formulae $\phi$ and $\psi$ such that $\phi \wedge \psi$ is
unsatisfiable, a \emph{Lyndon interpolant} is a formula $I$ such that
$\phi \models I$, the formula $I \wedge \psi$ is unsatisfiable,
$\voc{I} \subseteq \voc{\phi} \cap \voc{\psi}$, $\pset{+}{I} \subseteq
\pset{+}{\phi} \cap \pset{+}{\psi}$ and $\pset{-}{I} \subseteq
\pset{-}{\phi} \cap \pset{-}{\psi}$.
\end{definition}

In the rest of this section, let us fix an automaton $\A =
\tuple{\Sigma,X,Q,\iota,F,\Delta}$. Due to the above observation, none
of the interpolants considered will be control-local and we shall use
the term \emph{local} to denote time-local interpolants, with no free
variables.

\begin{definition}\label{def:generalized-lyndon-interpolant}
  Given a non-empty sequence of input events $\alpha = a_1 \ldots a_n
  \in \Sigma^*$, a \emph{generalized Lyndon interpolant (GLI)} is a
  sequence $(I_0,\ldots,I_n)$ of formulae such that, for all $k \in
  [n-1]$: \begin{compactenum}
    \item\label{it1:generalized-lyndon-interpolant} $\pset{-}{I_k} =
      \emptyset$,
    \item\label{it2:generalized-lyndon-interpolant} $\stamp{\iota}{0}
      \models I_0$ and \(I_k \wedge \Big(\bigwedge_{q(\vec{y})
        \arrow{a_i(X)}{} \psi \in \Delta} \forall y_1 \ldots \forall
      y_{\#(q)} ~.~ \stamp{q}{k}(\vec{y}) \rightarrow
      \stamp{\psi}{k+1}\Big) \models I_{k+1}\),
    \item\label{it3:generalized-lyndon-interpolant} $I_n \wedge
      \bigwedge_{q \in Q \setminus F} \forall y_1 \ldots \forall
      y_{\#(q)} ~.~ q(\vec{y})$ is unsatisfiable.
  \end{compactenum}
  Moreover, the GLI is \emph{local} if and only if $\voc{I_k}
  \subseteq \stamp{Q}{k}$, for all $k \in [n]$.
\end{definition}
The following proposition states the existence of local GLI for the
theories in which Lyndon's Interpolation Theorem holds.

\begin{proposition}\label{prop:local-gli}
  If there exists a Lyndon interpolant for any two formulae $\phi$ and
  $\psi$, such that $\phi \wedge \psi$ is unsatisfiable, then any
  sequence of input events $\alpha = a_1 \ldots a_n \in \Sigma^*$,
  such that $\accform{\alpha}$ is unsatisfiable, has a local GLI
  $(I_0,\ldots,I_n)$.
\end{proposition}
\proof{ By definition, $\accform{\alpha}$ is the formula:
  \[\stamp{\iota}{0} \wedge 
 \bigwedge_{i=1}^n \bigwedge_{q(\vec{y}) \arrow{a_i(X)}{} \psi \in
   \Delta} \forall y_1 \ldots \forall y_{\#(q)} ~.~
 \stamp{q}{i-1}(\vec{y}) \rightarrow \stamp{\psi}{i} \wedge
 \bigwedge_{q \in Q \setminus F} \forall y_1 \ldots \forall y_{\#(q)}
 ~.~ \stamp{q}{n}(\vec{y}) \rightarrow \bot\]
 We define the formulae: 
 \[\begin{array}{rcl}
 \varphi_i & \isdef & \bigwedge_{q(\vec{y}) \arrow{a_i(X)}{} \psi \in \Delta} \forall y_1 \ldots \forall y_{\#(q)} ~.~
 \stamp{q}{i-1}(\vec{y}) \rightarrow \stamp{\psi}{i},~ \text{for all}~ i \in [1,n] \\
 \psi & \isdef & \bigwedge_{q \in Q \setminus F} \forall y_1 \ldots \forall y_{\#(q)}
 ~.~ \stamp{q}{n}(\vec{y}) \rightarrow \bot
 \end{array}\]
 Observe that $\voc{\stamp{\iota}{0}} \subseteq \stamp{Q}{0}$,
 $\voc{\varphi_i} \subseteq
 \stamp{Q}{i-1}\cup\stamp{Q}{i}\cup\stamp{X}{i}$, for all $i \in
       [1,n]$, and $\voc{\psi} \subseteq \stamp{Q}{n}$.  We apply
       Lyndon's Interpolation Theorem for the formulae
       $\stamp{\iota}{0}$ and $\bigwedge_{i=1}^n \varphi_i \wedge
       \psi$ and obtain a formula $I_0$, such that $\stamp{\iota}{0}
       \models I_0$, $I_0 \wedge \bigwedge_{i=1}^n \varphi_i \wedge
       \psi$ is unsatisfiable, $\voc{I_0} \subseteq
       \voc{\stamp{\iota}{0}} \cap (\bigcup_{i=1}^n \voc{\varphi_i}
       \cup \voc{\psi}) \subseteq \stamp{Q}{0}$ and $\pset{-}{I_0}
       \subseteq \pset{-}{\stamp{\iota}{0}} \cap (\bigcup_{i=1}^n
       \pset{-}{\varphi} \cup \pset{-}{\psi}) = \emptyset$. Repeating
       the reasoning for the formulae $I_0 \wedge \varphi_1$ and
       $\bigwedge_{i=2}^n \varphi_i \wedge \psi$, we obtain $I_1$,
       such that $I_0 \wedge \varphi_1 \models I_1$, $I_1 \wedge
       \bigwedge_{i=2}^n \varphi_i \wedge \psi$ is unsatisfiable,
       $\voc{I_1} \subseteq (\voc{I_0} \cup \voc{\varphi_1}) \cap
       (\bigcup_{i=2}^n \voc{\varphi_i} \cup \voc{\psi}) \subseteq
       \stamp{Q}{1}$ and $\pset{-}{I_1} \subseteq (\pset{-}{I_0} \cup
       \pset{-}{\varphi_1}) \cap (\bigcup_{i=2}^n \pset{-}{\varphi_i}
       \cup \pset{-}{\psi}) = \emptyset$. Continuing in this way, we
       obtain formulae $I_0, I_1, \ldots, I_{n}$ as required. \qed}

The main problem with the local GLI construction described in the
proof of Proposition \ref{prop:local-gli} is that the existence of
Lyndon interpolants (Definition \ref{def:lyndon-interpolant}) is
guaranteed in principle, but the proof is non-constructive. Building
an interpolant for an unsatisfiable conjunction of formulae $\phi
\wedge \psi$ is typically the job of the decision procedure that
proves the unsatisfiability and, in general, there is no such
procedure, when $\phi$ and $\psi$ contain predicates and have
non-trivial quantifier alternation. In this case, some provers use
instantiation heuristics for the universal quantifiers that are
sufficient for proving unsatisfiability, however these heuristics are
not always suitable for interpolant generation. Consequently, from now
on, we assume the existence of an effective Lyndon interpolation
procedure only for decidable theories, such as the quantifier-free
linear (integer) arithmetic with uninterpreted functions (UFLIA,
UFLRA, etc.) \cite{RybalchenkoSofronieStokkermans}. 

This is where the predicate-free path formulae (Definition
\ref{def:substform}) come into play. For a given event sequence
$\alpha$, the automaton $\A$ accepts a word $w$ such that $\event{w} =
\alpha$ if and only if $\substform{\alpha}$ is satisfiable. Assuming
further that the equality atoms in the transition rules of $\A$ are
written in the language of a decidable first order theory, such as
Presburger arithmetic, Lemma \ref{lemma:quant-pred-acceptance} gives
us an effective way of checking emptiness of $\A$, relative to a given
event sequence. However, this method does not cope well with lazy
annotation, because there is no way to extract, from the
unsatisfiability proof of $\substform{\alpha}$, the interpolants
needed to annotate $\alpha$. This is because \begin{inparaenum}[(i)]
\item\label{it1:gli} the formula $\substform{\alpha}$, obtained by
  repeated substitutions (Definition \ref{def:substform}) loses track
  of the steps of the execution, and
\item\label{it2:gli} quantifiers that occur nested in
  $\substform{\alpha}$ make it difficult to write $\substform{\alpha}$
  as an unsatisfiable conjunction of formulae from which interpolants
  are extracted (Definition \ref{def:lyndon-interpolant}).
\end{inparaenum}

The solution we adopt for the first issue (\ref{it1:gli}) consists in
partially recovering the time-stamped structure of the acceptance
formula $\accform{\alpha}$ using the formula $\quantform{\alpha}$, in
which only transition quantifiers occur. The second issue
(\ref{it2:gli}) is solved under the additional assuption that the
theory of the data domain $\Data$ has \emph{witness-producing
  quantifier elimination}. More precisely, we assume that, for each
formula $\exists x ~.~ \phi(x)$, there exists an effectively
computable term $\tau$, in which $x$ does not occur, such that
$\exists x ~.~ \phi$ and $\phi[\tau/x]$ are equisatisfiable. These
terms, called \emph{witness terms} in the following, are actual
definitions of the Skolem function symbols from the following folklore
theorem:

\begin{theorem}[\cite{BorgerGraedelGurevich97}]\label{thm:skolem}
  Given $Q_1 x_1 \ldots Q_n x_n ~.~ \phi$ a first order sentence,
  where $Q_1, \ldots, Q_n \in \set{\exists,\forall}$ and $\phi$ is
  quantifier-free, let $\eta_i \isdef f_i(y_1,\ldots,y_{k_i})$ if $Q_i
  = \forall$ and $\eta_i \isdef x_i$ if $Q_i = \exists$, where $f_i$
  is a fresh function symbol and $\set{y_1, \ldots, y_{k_i}} =
  \set{x_j \mid j < i,~ Q_j = \exists}$. Then the entailment \(Q_1 x_1
  \ldots Q_n x_n ~.~ \phi \models \phi[\eta_1/x_1,\ldots,\eta_n/x_n]\)
  holds.
\end{theorem}
\proof{See \cite[Theorem 2.1.8]{BorgerGraedelGurevich97} and
  \cite[Lemma 2.1.9]{BorgerGraedelGurevich97}. \qed}

Examples of witness-producing quantifier elimination procedures can be
found in the literature for e.g. linear integer (real) arithmetic
(LIA,LRA), Presburger arithmetic and boolean algebra of sets and
Presburger cardinality constraints (BAPA)
\cite{KuncakMayerPiskacSuter12}.  


Under the assumption that witness terms can be effectively built, let
us describe the generation of a non-local GLI for a given input event
sequence $\alpha = a_1 \ldots a_n$. First, we generate successively
the acceptance formula $\accform{\alpha}$ and its equisatisfiable
forms $\quantform{\alpha} = Q_1x_1 \ldots Q_mx_m ~.~ \widehat{\Phi}$
and $\substform{\alpha} = Q_1x_1 \ldots Q_mx_m ~.~ \overline{\Phi}$,
both written in prenex form, with matrices $\widehat{\Phi}$ and
$\overline{\Phi}$, respectively. Because we assumed that the
first order theory of $\Data$ has quantifier elimination, the
satisfiability problem for $\substform{\alpha}$ is decidable. If
$\substform{\alpha}$ is satisfiable, we build a counterexample for
emptiness $w$ such that $\event{w}=\alpha$ and $\data{w}$ is a
satisfying assignment for $\substform{\alpha}$. Otherwise,
$\substform{\alpha}$ is unsatisfiable and there exist witness terms
$\tau_{i_1} \ldots \tau_{i_\ell}$, where $\set{i_1, \ldots, i_\ell} =
\set{j \in [1,m] \mid Q_j = \forall}$, such that
$\overline{\Phi}[\tau_{i_1}/x_{i_1}, \ldots,
  \tau_{i_\ell}/x_{i_\ell}]$ is unsatisfiable (Theorem
\ref{thm:skolem}). Then it turns out that the formula
$\widehat{\Phi}[\tau_{i_1}/x_{i_1}, \ldots,
  \tau_{i_\ell}/x_{i_\ell}]$, obtained analogously from the matrix of
$\quantform{\alpha}$, is unsatisfiable as well (Lemma
\ref{lemma:subst-quant}). Because this latter formula is structured as
a conjunction of formulae $\stamp{\iota}{0} \wedge \phi_1 \ldots
\wedge \phi_n \wedge \psi$, where $\voc{\phi_k} \cap \stamp{Q}{\leq n}
\subseteq \stamp{Q}{k-1} \cup \stamp{Q}{k}$ and $\voc{\psi} \cap
\stamp{Q}{\leq n} \subseteq \stamp{Q}{n}$, it is now possible to use
an existing interpolation procedure for the quantifier-free theory of
$\Data$, extended with uninterpreted function symbols, to compute a
sequence of non-local GLI $(I_0, \ldots, I_n)$ such that $\voc{I_k}
\cap \stamp{Q}{\leq n} \subseteq \stamp{Q}{k}$, for all $k \in [n]$.


\begin{example}[Contd. from Examples \ref{ex:quant-elim} and \ref{ex:pred-elim}]\label{ex:gli}
The formula $\substform{\alpha}$ (Example \ref{ex:pred-elim}) is
unsatisfiable and let $\tau_2 = z_1$ be the witness term for the
universally quantified variable $z_2$. Replacing $z_2$ with $\tau_2$
in the matrix of $\quantform{\alpha}$ (Example \ref{ex:quant-elim})
yields the unsatisfiable conjunction: 
\vspace*{-\baselineskip}
\[\begin{array}{c}
z_1 \geq 0 \wedge \stamp{q}{0}(z_1)  ~\wedge~
\stamp{q}{0}(z_1) \rightarrow \stamp{x}{1} \geq 0 \wedge (z_1 \geq z_1 \rightarrow \stamp{q}{1}(\stamp{x}{1}+z_1)) ~\wedge~ \\
\stamp{q}{1}(\stamp{x}{1}+z_1) \rightarrow \stamp{x}{1}+z_1 < 0 \wedge \stamp{q_f}{2}(\stamp{x}{2}+\stamp{x}{1}+z_1)
\end{array}\]
A non-local GLI for the above is $(\stamp{q}{0}(z_1) \wedge
z_1\geq 0,~ \stamp{x}{1} \geq 0 \wedge \stamp{q}{1}(\stamp{x}{1}+z_1)
\wedge z_1\geq0,~\bot)$. \hfill$\blacksquare$
\end{example}

A function $\xi : \nat \rightarrow \nat$ is [strictly]
\emph{monotonic} iff for each $n < m$ we have $\xi(n) \leq \xi(m)$
     [$\xi(n) < \xi(m)$] and \emph{finite-range} iff for each $n \in
     \nat$ the set $\set{m \mid \xi(m)=n}$ is finite. If $\xi$ is
     finite-range, we denote by $\xi_{\max}^{-1}(n) \in \nat$ the
     maximal value $m$ such that $\xi(m)=n$. The lemma below gives the
     proof of correctness for the construction of non-local GLI.

\begin{lemma}\label{lemma:subst-quant}
  Given a non-empty input event sequence $\alpha = a_1 \ldots a_n \in
  \Sigma^*$, such that $\accform{\alpha}$ is unsatisfiable, let
  $Q_1x_1 \ldots Q_mx_m ~.~ \widehat{\Phi}$ be a prenex form of
  $\quantform{\alpha}$ and let $\xi : [1,m] \rightarrow [n]$ be a
  monotonic function mapping each transition quantifier to the minimal
  index from the sequence $\quantformn{\alpha_0}, \ldots,
  \quantformn{\alpha_n}$ where it occurs. Then one can effectively build:
  \begin{compactenum}
  \item\label{it1:subst-quant} witness terms $\tau_{i_1}, \ldots,
    \tau_{i_\ell}$, where \(\set{i_1, \ldots,i_\ell} = \set{j \in
      [1,m] \mid Q_j = \forall}\) and \(\voc{\tau_{i_j}} \subseteq
    \stamp{X}{\leq \xi(i_j)} \cup \set{x_k \mid k < i_j, Q_k =
      \exists},~ \forall j \in [1,\ell]\) such that
    $\widehat{\Phi}[\tau_{i_1}/x_{i_1}, \ldots,
      \tau_{i_\ell}/x_{i_\ell}]$ is unsatisfiable, and
  \item\label{it2:subst-quant} a GLI $(I_0, \ldots, I_n)$ for
    $\alpha$, such that \(\voc{I_k} \subseteq \stamp{Q}{k} \cup
    \stamp{X}{\leq k} \cup \set{x_j \mid j < \xi^{-1}(k),~ Q_j =
      \exists}\), for all $k \in [n]$. 
  \end{compactenum}
\end{lemma}
\proof{(\ref{it1:subst-quant}) If $\accform{\alpha}$ is unsatisfiable,
  by Lemmas \ref{lemma:quant} and \ref{lemma:subst}, we obtain that,
  successively $\quantform{\alpha}$ and $\substform{\alpha}$ are
  unsatisfiable. Let $Q_1x_1 \ldots Q_mx_m ~.~ \widehat{\Phi}$ and
  $Q_1x_1 \ldots Q_mx_m ~.~ \overline{\Phi}$ be prenex forms for
  $\quantform{\alpha}$ and $\substform{\alpha}$, respectively. Since
  we assumed that the first order theory of the data domain has
  witness-producing quantifier elimination, using Theorem
  \ref{thm:skolem} one can effectively build witness terms
  $\tau_{i_1}, \ldots, \tau_{i_\ell}$, where
  $\set{i_1,\ldots,i_\ell}=\set{i \in [1,m] \mid Q_i=\forall}$
  and: \begin{compactitem}
  \item $\voc{\tau_{i_j}} \subseteq \stamp{X}{\leq \xi(i_j)} \cup
    \set{x_k \mid k < i_j, Q_k = \exists}$, for all $j \in [1,\ell]$ and
  \item $\overline{\Phi}[\tau_{i_1}/x_{i_1}, \ldots,
    \tau_{i_\ell}/x_{i_\ell}]$ is unsatisfiable.
  \end{compactitem}
  Let $\widehat{\Phi}_0$, \ldots, $\widehat{\Phi}_n$ be the sequence
  of quantifier-free formulae, defined as follows: \begin{compactitem}
  \item $\widehat{\Phi}_0$ is the matrix of some prenex form of
    $\stamp{\iota}{0}$,
  \item for all $i=1,\ldots,n$, let $\widehat{\Phi}_i$ be the matrix
    of some prenex form of:
    \[\widehat{\Phi}_i \isdef \widehat{\Phi}_{i-1} \wedge
     \underbrace{\bigwedge_{\begin{array}{l}
         \scriptstyle{\stamp{q}{i-1}(t_1,\ldots,t_{\#(q)}) \text{ occurs in } \widehat{\Phi}_{i-1}} \\
        \scriptstyle{q(y_1,\ldots,y_{\#(q)}) \arrow{a_i(X)}{} \psi \in \Delta} 
     \end{array}} \stamp{q}{i-1}(t_1,\ldots,t_{\#(q)}) \rightarrow 
     \stamp{\psi}{i}[t_1/y_1, \ldots, t_{\#(q)}/y_{\#(q)}]}_{\isdef \phi_i}\]  
  \end{compactitem}
  It is easy to see that $\widehat{\Phi}$ is the matrix of some prenex
  form of:
  \[\widehat{\Phi}_n \wedge \underbrace{\bigwedge_{\begin{array}{l}
        \scriptstyle{\stamp{q}{n}(t_1,\ldots,t_{\#(q)}) \text{ occurs in } \widehat{\Phi}_n} \\
        \scriptstyle{q \in Q \setminus F}
    \end{array}} \stamp{q}{n}(t_1,\ldots,t_{\#(q)}) \rightarrow \bot}_{\isdef \psi}\]
  Applying the equivalence from Fact \ref{fact:subst} in the proof of
  Lemma \ref{lemma:subst}, we obtain a sequence of quantifier-free
  formulae $\overline{\Phi}_0, \ldots, \overline{\Phi}_n$ such that
  $\overline{\Phi}_i \equiv \widehat{\Phi}_i$, for all $i \in [n]$ and
  $\overline{\Phi}$ is obtained from $\overline{\Phi}_n$ by replacing
  each occurrence of a predicate atom $q(t_1,\ldots,t_{\#(q)})$ in
  $\overline{\Phi}_n$ by $\bot$ if $q \in Q\setminus F$ and by $\top$
  if $q \in F$. Clearly $\overline{\Phi} \equiv \widehat{\Phi}$, thus
  $\widehat{\Phi}[\tau_{i_1}/x_{i_1}, \ldots,
    \tau_{i_\ell}/x_{i_\ell}] \equiv
  \overline{\Phi}[\tau_{i_1}/x_{i_1}, \ldots,
    \tau_{i_\ell}/x_{i_\ell}] \equiv \bot$.

  \noindent(\ref{it2:subst-quant}) With the notation introduced at
  point (\ref{it1:subst-quant}), we have $\widehat{\Phi} =
  \widehat{\Phi}_0 \wedge \bigwedge_{i=1}^n \phi_i \wedge
  \psi$. Consider the sequence of witness terms $\tau_{i_1}, \ldots,
  \tau_{i_\ell}$, whose existence is proved by point
  (\ref{it1:subst-quant}). Because $\voc{\tau_{i_j}} \subseteq
  \stamp{X}{\leq \xi(i_j)} \cup \set{x_k \mid k < i_j, Q_k =
    \exists}$, for all $j \in [1,\ell]$, and moreover $\xi^{-1}$ is
  strictly monotonic, we obtain: \begin{compactitem}
  \item $\voc{\widehat{\Phi}_0[\tau_{i_1}/x_{i_1}, \ldots,
      \tau_{i_\ell}/x_{i_\ell}]} \subseteq \stamp{Q}{0} \cup
    \stamp{X}{0} \cup \set{x_j \mid j < \xi_{\max}^{-1}(0), Q_j = \exists}$,
  \item $\voc{\phi_i[\tau_{i_1}/x_{i_1}, \ldots,
      \tau_{i_\ell}/x_{i_\ell}]} \subseteq \stamp{Q}{i-1} \cup
    \stamp{Q}{i} \cup \stamp{X}{\leq i} \cup \set{x_j \mid j <
      \xi_{\max}^{-1}(i), Q_j = \exists}$, for all $i \in [1,n]$,
  \item $\voc{\psi[\tau_{i_1}/x_{i_1}, \ldots,
      \tau_{i_\ell}/x_{i_\ell}]} \subseteq \stamp{Q}{n} \cup
    \stamp{X}{\leq n} \cup \set{x_j \mid j \in [1,m], Q_j = \exists}$.
  \end{compactitem}
  By repeatedly applying Lyndon's Interpolation Theorem, we obtain a
  sequence of formulae $(I_0, \ldots, I_n)$ such that: \begin{compactitem}
    \item $\widehat{\Phi}_0[\tau_{i_1}/x_{i_1}, \ldots,
      \tau_{i_\ell}/x_{i_\ell}] \models I_0$ and $\voc{I_0} \subseteq
      \stamp{Q}{0} \cup \stamp{X}{0} \cup \set{x_j \mid j <
        \xi_{\max}^{-1}(0), Q_j = \exists}$,
    \item $I_{k-1} \wedge \phi_i[\tau_{i_1}/x_{i_1}, \ldots,
      \tau_{i_\ell}/x_{i_\ell}] \models I_k$ and $\voc{I_k} \subseteq
      \stamp{Q}{k} \cup \stamp{X}{\leq k} \cup \set{x_j \mid j <
       \xi_{\max}^{-1}(k), Q_j = \exists}$, for all $k \in [1,n]$,
    \item $I_n \wedge \psi[\tau_{i_1}/x_{i_1}, \ldots,
      \tau_{i_\ell}/x_{i_\ell}]$ is unsatisfiable.
  \end{compactitem}
  To show that $(I_0,\ldots,I_n)$ is a GLI for $a_1\ldots a_n$, it is
  sufficient to notice that \[\bigwedge_{q(\vec{y}) \arrow{a_k(X)}{}
    \psi \in \Delta} \forall y_1 \ldots \forall y_{\#(q)} ~.~
  \stamp{q}{k}(\vec{y}) \rightarrow \stamp{\psi}{k+1} \models \phi_k\]
  for all $k \in [1,n]$. Consequently, we obtain: \begin{compactitem}
    \item $\stamp{\iota}{0} \models \widehat{\Phi}_0 \models I_0$, by
      Theorem \ref{thm:skolem},
    \item $I_{k-1} \wedge \Big(\bigwedge_{q(\vec{y}) \arrow{a_k(X)}{}
      \psi \in \Delta} \forall y_1 \ldots \forall y_{\#(q)} ~.~
      \stamp{q}{k-1}(\vec{y}) \rightarrow \stamp{\psi}{k}\Big) \models
      I_{k-1} \wedge \phi_k \models I_k$, and 
    \item $I_n \wedge \Big(\bigwedge_{q \in Q \setminus F} \forall y_1
      \ldots \forall y_{\#(q)} ~.~ q(\vec{y}) \rightarrow \bot\Big)
      \models I_n \wedge \psi \models \bot$,     
  \end{compactitem}
  as required by Definition
  \ref{def:generalized-lyndon-interpolant}. \qed}

In conclusion, under two assumptions about the first order theory of
the data domain, namely the\begin{inparaenum}[(i)]
\item witness-producing quantifier elimination, and
\item Lyndon interpolation for the quantifier-free fragment with
  uninterpreted functions,
\end{inparaenum}
we developped a rather generic method that produces generalized Lyndon
interpolants for unfeasible input event sequences. Moreover, each
formula $I_k$ in the interpolant refers only to the current predicate
symbols $\stamp{Q}{I_k}$, the current and past input variables
$\stamp{X}{\leq k}$ and the existentially quantified transition
variables introduced at the previous steps $\set{x_j \mid j <
  \xi_{\max}^{-1}(k), Q_j = \exists}$. The remaining question is how
to use such non-local interpolants to label the unfolding of an
automaton (Definition \ref{def:unfolding}) and to compute the coverage
between nodes of the unfolding (Definition \ref{def:coverage}).

\subsection{Unfolding with Non-local Interpolants}

As required by Definition \ref{def:unfolding}, the unfolding $U$ of an
automaton $\A = \tuple{\Sigma,X,Q,\iota,F,\Delta}$ is labeled by
formulae $U(\alpha) \in \posforms(Q,\emptyset)$, with no free symbols,
other than predicate symbols, such that the labeling is compatible
with the transition relation of the automaton, according to the point
(\ref{it3:unfolding}) of Definition \ref{def:unfolding}. The following
lemma describes the refinement of the labeling of an input sequence
$\alpha$ of length $n$ by a non-local GLI $(I_0, \ldots, I_n)$, such
that $\voc{I_k} \subseteq \stamp{Q}{k} \cup \stamp{X}{\leq k} \cup
\vec{x}_k$, where $\vec{x}_k$ are the existentially quantified
variables from the prenex normal form of $\quantform{\alpha_k}$. 

\begin{lemma}\label{lemma:refinement}
  Let $U$ be an unfolding of an automaton $\A =
  \tuple{\Sigma,X,Q,\iota,F,\Delta}$ such that $\alpha = a_1\ldots a_n
  \in \dom(U)$ and $(I_0, \ldots, I_n)$ be a GLI for $\alpha$. The
  mapping $U' : \dom(U) \rightarrow \posforms(Q,\emptyset)$ defined as: 
  \begin{compactitem}
  \item $U'(\alpha_k) = U(\alpha_k) \wedge J_k$, for all $k \in [n]$,
    where $J_k$ is the formula obtained from $I_k$ by replacing each
    time-stamped predicate symbol $\stamp{q}{k}$ by $q$ and
    existentially quantifying each free variable in $I_k$,
\item $U'(\beta) = U(\beta)$ if $\beta \in \dom(U)$ and $\beta
  \not\prefix \alpha$,
  \end{compactitem}
  is an unfolding of $\A$. 
\end{lemma}
\proof{
  The new set of formulae $U'(\alpha_0), \ldots, U'(\alpha_n)$ complies
  with Definition \ref{def:unfolding}, because: \begin{compactitem}
  \item $U'(\alpha_0) \equiv \iota$, since, by point
    \ref{it2:generalized-lyndon-interpolant} of Definition
    \ref{def:generalized-lyndon-interpolant}, we have $\stamp{\iota}{0}
    \models I_0$, thus $\iota \models J_0$ and $U'(\alpha_0) =
    U(\alpha_0) \wedge J_0 \equiv \iota \wedge J_0 \equiv \iota$, and
  \item by Definition \ref{def:generalized-lyndon-interpolant}
    (\ref{it3:generalized-lyndon-interpolant}) we have, for all $k \in
    [n-1]$:
    \[I_k \wedge \bigwedge_{\scriptscriptstyle{q(\vec{y}) \arrow{a_k(X)}{} \psi \in \Delta}} 
    \forall y_1 \ldots \forall y_{\#(q)} ~.~ \stamp{q}{k}(\vec{y})
    \rightarrow \stamp{\psi}{k+1} \models I_{k+1}\] We write
    $\overstamp{I}{j}_k$ for the formula in which each predicate
    symbol $\stamp{q}{k}$ is replaced by $\stamp{q}{j}$. Then the
    following entailment holds:
    \[\overstamp{I}{0}_k \wedge \bigwedge_{\scriptscriptstyle{q(\vec{y}) \arrow{a_k(X)}{} \psi \in \Delta}} 
    \forall y_1 \ldots \forall y_{\#(q)} ~.~ \stamp{q}{0}(\vec{y})
    \rightarrow \stamp{\psi}{1} \models \overstamp{I}{1}_{k+1}\]
    Because $J_k$ is obtained by removing the time stamps from the predicate symbols and  
    existentially quantifying all the free variables of $I_k$, we also
    obtain, applying Fact \ref{fact:ex-entail} below:
    \[\stamp{J}{0}_k \wedge \bigwedge_{\scriptscriptstyle{q(\vec{y}) \arrow{a_k(X)}{} \psi \in \Delta}} 
    \forall y_1 \ldots \forall y_{\#(q)} ~.~ \stamp{q}{0}(\vec{y})
    \rightarrow \stamp{\psi}{1} \models \stamp{J}{1}_{k+1}\] Since $U$
    satisfies the labeling condition of Definition \ref{def:unfolding}
    (\ref{it3:unfolding}) and $U'(\alpha_k) = U(\alpha_k) \wedge J_k$,
    we obtain, as required: 
    \[\stamp{U'(\alpha_k)}{0} \wedge \bigwedge_{\scriptscriptstyle{q(\vec{y}) \arrow{a_k(X)}{} \psi \in \Delta}} 
    \forall y_1 \ldots \forall y_{\#(q)} ~.~ \stamp{q}{0}(\vec{y})
    \rightarrow \stamp{\psi}{1} \models
    \stamp{U'(\alpha_{k+1})}{1}\enspace.\]
  \end{compactitem}  

  \begin{fact}\label{fact:ex-entail}
    Given formulae $\phi(\vec{x}, \vec{y})$ and $\psi(\vec{x})$ such
    that $\phi(\vec{x}, \vec{y}) \models \psi(\vec{x})$, we also have
    $\exists \vec{x} ~.~ \phi(\vec{x},\vec{y}) \models \exists \vec{x}
    ~.~ \psi(\vec{x})$. 
  \end{fact}
  \proof{ For each choice of a valuation for the existentially
    quantified variables on the left-hand side, we chose the same
    valuation for the variables on the right-hand side. \qed}\qed}

Observe that, by Lemma \ref{lemma:subst-quant}
(\ref{it2:subst-quant}), the set of free variables of a GLI formula
$I_k$ consists of \begin{inparaenum}[(i)]
\item variables $\stamp{X}{\leq k}$ keeping track of data values seen
  in the input at some earlier moment in time, and
\item variables that track past choices made within the transition
  rules.
\end{inparaenum}
Basically, it is not important when exactly in the past a certain
input has been read or when a choice has been made, as only the value
of the variable determines the future behavior. Intuitively,
existential quantification of these variables does the job of ignoring
when in the past these values have been seen.

The last ingredient of the lazy annotation semi-algorithm based on
unfoldings consist in the implementation of the coverage check, when
the unfolding of an automaton is labeled with conjunctions of
existentially quantified formulae with predicate symbols, obtained
from interpolation. By Definition \ref{def:coverage}, checking whether
a given node $\alpha \in \dom(U)$ is covered amounts to finding a
prefix $\alpha' \prefix \alpha$ and a node $\beta \in \dom(U)$ such
that $U(\alpha') \models U(\beta)$, or equivalently, the formula
$U(\alpha') \wedge \neg U(\beta)$ is unsatisfiable. However, the
latter formula, in prenex form, has quantifier prefix in the language
$\exists^*\forall^*$ and, as previously mentioned, the satisfiability
problem for such formulae becomes undecidable when the data theory
subsumes Presburger arithmetic \cite{Halpern91}.

Nevertheless, if we require just a yes/no answer (i.e.\ not an
interpolant) recently developped quantifier instantiation heuristics
\cite{ReynoldsKK17} perform rather well in answering a large number of
queries in this class. Observe, moreover, that coverage does not need
to rely on a complete decision procedure. If the prover fails in
answering the above satisfiability query, then the semi-algorithm
assumes that the node is not covered and continues exploring its
successors. Failure to compute complete coverage may lead to
divergence (non-termination) and ultimately, to failure to prove
emptiness, but does not affect the soundness of the semi-algorithm
(real counterexamples will still be found).

\ifLongVersion
\section{Applications}
\label{sec:applications}

The main application of first order alternating automata is checking
inclusion between various classes of automata extended with variables
ranging over infinite domains that recognize languages over infinite
alphabets. The most widely known such classes are \emph{timed
  automata} \cite{AlurDill94} and \emph{finite-memory (register)
  automata} \cite{KaminskiFrancez94}. In both cases, complementation
is not possible inside the class and inclusion is undecidable. Our
contribution is providing a systematic semi-algorithm for these
decision problems. In addition, the method described in
\S\ref{sec:emptiness} can extend our previous \emph{generic register
  automata} \cite{IosifRV16} inclusion checking framework, by allowing
monitor (right-hand side) automata to have local variables, that are
not visible in the language.

Another application is checking safety (mutual exclusion, absence of
deadlocks, etc.) and liveness (termination, lack of starvation, etc.)
properties of parameterized concurrent programs, consisting of an
unbounded number of replicated threads that communicate via a fixed
set of global variables (locks, counters, etc.). The verification of
parametric programs has been reduced to checking the emptiness of a
(possibly infinite) sequence of first order alternating automata,
called \emph{predicate automata} \cite{Farzan15,Farzan16}, encoding
the inclusion of the set of traces of a parametric concurrent program
into increasingly general \emph{proof spaces}, obtained by
generalization of counterexamples. The program and the proof spaces
are first order alternating automata over the infinite alphabet of
pairs consisting of program statements and thread identifiers.

\subsection{Timed Automata}
\label{sec:ta}

The standard definition of a finite \emph{timed word} is a sequence of
pairs $(a_1,\tau_1), \ldots, (a_n,\tau_n) \in (\Sigma \times
\real)^*$, where $\real$ is the set of real numbers, such that $0 \leq
\tau_i < \tau_{i+1}$, for all $i \in [1,n-1]$. Intuitively, $\tau_i$
is the moment in time where the input event $a_i$ occurs. Given a set
$C$ of \emph{clocks}, the set $\Phi(C)$ of \emph{clock constraints} is
defined inductively as the set of formulae $x \leq c$, $x \geq c$,
$\neg\delta$, $\delta_1 \wedge \delta_2$, where $x \in C$, $c \in
\rat$ is a rational constant and $\delta,\delta_1,\delta_2 \in
\Phi(X)$. 

A \emph{timed automaton} is a tuple $T = \tuple{\Sigma,S,S_0,F,C,E}$,
where: $\Sigma$ is a finite set of input events, $S$ is a finite set
of states, $S_0, F \subseteq S$ are sets of initial and final states,
respectively, $C$ is a finite set of clocks and $E \subseteq S \times
\Sigma \times S \times 2^C \times \Phi(C)$ is the set of transitions
$(s,a,s',\lambda,\delta)$ from state $s$ to state $s'$ with symbol
$a$, $\lambda$ is the set of clocks to be reset and $\delta$ is a
clock constraint. A run of $T$ over a timed word $w = (a_1,\tau_1)
\ldots (a_n,\tau_n)$ is a sequence $(s_0,\gamma_0) \ldots
(s_n,\gamma_n)$, where $s_i \in S$, $\gamma_i : C \rightarrow \real$
are clock valuations, for all $i \in [n]$, and: \begin{compactitem}
\item $s_0 \in S_0$ and $\gamma_0(x)=0$ for all $x \in C$, 
\item for all $i \in [n]$, there exists a transition
  $(s_i,a_i,s_{i+1},\lambda_i,\delta_i) \in E$ such that $\gamma_i +
  \tau_{i+1} - \tau_i \models \delta_i$, and for all $x \in C$,
  $\gamma_{i+1}(x) = 0$ if $x \in \lambda_i$ and $\gamma_{i+1}(x) = \gamma_i(x)
  + \tau_{i+1}-\tau_i$, otherwise. Here $\tau_0 \isdef 0$ and $\gamma_i +
  \tau_{i+1} - \tau_i$ is the valuation mapping each $x \in C$ to
  $\gamma_i(x) + \tau_{i+1} - \tau_i$.
\end{compactitem}
The run is accepting iff $s_n \in F$, in which case $T$ accepts
$w$. As usual, we denote by $\lang{T}$ the set of finite words
accepted by $T$. It is well-known that, in general, there is no timed
automaton accepting the complement language $(\Sigma\times\real)^*
\setminus \lang{T}$ and, moreover, the language inclusion problem is
undecidable \cite{AlurDill94}.

Given a timed automaton $T = \tuple{\Sigma,S,S_0,F,C,E}$, we define a
first order alternating automaton $\A_T =
\tuple{\Sigma,\set{t},Q_T,\iota_T,F_T,\Delta_T}$, with a single input
variable $t$, ranging over $\real$, such that each timed word $w =
(a_1,\tau_1) \ldots (a_n,\tau_n)$ corresponds to a unique data word
$d(w) = (a_1,\nu_1) \ldots (a_n,\nu_n)$ such that $\nu_i(t) = \tau_i$
for all $i \in [1,n]$ and $\lang{\A_T} = \set{d(w) \mid
  w\in\lang{T}}$. The only difficulty here is capturing the fact that
all the clocks of $T$ evolve at the same pace, which is easily done
using a technique from \cite{Fribourg98}, which replaces
each clock $x_i$ of $T$ by a variable $y_i$ tracking the difference
between the values of $t$ and $x_i$.

Formally, if $C = \set{x_1,\ldots,x_k}$ and $S =
\set{s_1,\ldots,s_m}$, we define $Q_T \isdef \set{q_1, \ldots, q_m}$ ,
where $\#(q_i) = k+1$ for all $i \in [1,m]$, $\iota_T \isdef
\bigvee_{s_i \in S_0} q_i(0,\ldots,0)$, $F_T \isdef \set{q_i \mid s_i
  \in F}$ and, for each transition $(s_i,a,s_j,\lambda,\delta) \in E$,
$\Delta_T$ contains the rule:
\[q_i(y_1,\ldots,y_k,z) \arrow{a(t)}{} t > z \wedge \delta(z-y_1,\ldots,z-y_k) \wedge q_j(y'_1,\ldots,y'_k,t)\]
where $y'_i$ stands for $z$ if $x_i \in \lambda$ and for $y_i$, 
otherwise.  Moreover, nothing else is in $\Delta_T$. We establish the
following connection between a timed automaton and its corresponding
first order alternating automaton.

\begin{proposition}\label{prop:timed-alternating}
  Given a timed automaton $T = \tuple{\Sigma,S,S_0,F,C,E}$, the
  first order alternating automaton $\A_T =
  \tuple{\Sigma,\set{t},Q_T,\iota_T,F_T,\Delta_T}$ recognizes the
  language $\lang{\A_T} = \set{d(w) \mid w \in \lang{T}}$. 
\end{proposition}
\proof{ ``$\subseteq$'' Let $w = (a_1,\nu_1) \ldots (a_n,\nu_n) \in
  \lang{\A_T}$ be a data word. We show the existence of a timed word
  $(a_1,\tau_1) \ldots (a_n,\tau_n) \in \lang{T}$ such that
  $\nu_i(t)=\tau_i$, for all $i \in [1,n]$, by induction on
  $n\geq0$. In fact we shall prove the following stronger
  statements: \begin{compactenum}
  \item each execution of $\A_T$ over $w$ starting with a cube $c \in
    \cube{\minsem{\iota_T}}$ is a linear tree, in which each node has
    at most one child.
  \item for each execution $q_{i_0}(d^0_1,\ldots,d^0_k,\tau_0) \ldots
    q_{i_n}(d^n_1,\ldots,d^n_k,\tau_n)$ of $\A_T$, $T$ has an
    execution $(s_{i_0}, \gamma_0) \ldots (s_{i_n}, \gamma_n)$ over
    the timed word $(a_1, \tau_1) \ldots (a_n, \tau_n)$, such that,
    for all $i \in [1,n]$ and all $\ell \in [1,k]$, we have
    $\gamma_i(x_\ell) = \tau_{i-1} - d^i_\ell$.
  \end{compactenum}
  The first point above is by inspection of $\iota_T = \bigvee_{s_i
    \in S_0} q_i(0,\ldots,0)$ and of the rules from
  $\Delta_T$. Indeed, each minimal model of $\iota_T$ corresponds to a
  cube $q(0,\ldots,0)$ and each rule has exactly one predicate atom on
  its right-hand side, thus each node of the execution will have at
  most one successor. The second point is by induction on $n\geq0$. 

  ``$\supseteq$'' Let $w = (a_1,\tau_1) \ldots (a_n,\tau_n) \in
  \lang{T}$ be a timed word. By induction on $n\geq0$, we show that
  for each run $(s_{i_0}, \gamma_0) \ldots (s_{i_n}, \gamma_n)$ of $T$
  over $w$, $\A_T$ has a linear execution
  $q_{i_0}(d^0_1,\ldots,d^0_k,\tau_0) \ldots
  q_{i_n}(d^n_1,\ldots,d^n_k,\tau_n)$ such that, for all $i \in [1,n]$
  and all $\ell \in [1,k]$, we have $\gamma_i(x_\ell) = \tau_{i-1} -
  d^i_\ell$. \qed}

An easy consequence is that the timed language inclusion problem
``given timed automata $T_1$ and $T_2$, does $\lang{T_1} \subseteq
\lang{T_2}$ ?'' is reduced in polynomial time to the emptiness problem
$\lang{\A_{T_1}} \cap \lang{\overline{\A_{T_2}}} = \emptyset$, for
which (\S\ref{sec:emptiness}) provides a semi-algorithm. Observe,
moreover, that no transition quantifiers are needed to encode timed
automata as first order alternating automata.

\subsection{Register Automata}
\label{sec:ra}

Finite-memory automata, most commonly referred to as register automata
\cite{KaminskiFrancez94} are among the first attempts at lifting the
finite alphabet restriction of classical Rabin-Scott automata. In a
nutshell, a register automaton is a finite-state automaton equipped
with a finite set of registers $x_1,\ldots,x_r$ able to copy input
values and compare them with subsequent input. Consequently, basic
results from classical automata theory, such as the pumping lemma or
the closure under complement do not hold in this model and, moreover,
inclusion of languages recognized by register automata is undecidable
\cite{NevenSchwentickVianu03}.

Let $\Sigma$ be an infinite alphabet, $\#$ be a symbol not in $\Sigma$
and $r>0$ be an integer constant, denoting the number of registers. An
\emph{assignment} is a word $\vec{v} = v_1 \ldots v_r$ such that if
$v_i = v_j$ and $i \neq j$ then $v_i = \#$, for all $i,j \in
     [1,r]$. We write $[\vec{v}]$ for the set $\set{v_i \mid i \in
       [1,r]}$ of values in the assignment $\vec{v}$. A
     \emph{finite-memory (register) automaton} is a tuple $R =
     \tuple{S,q_0,\vec{u}, \rho, \mu, F}$, where $S$ is a finite set
     of states, $q_0 \in S$ is the initial state, $\vec{u} = u_1
     \ldots u_r$ is the initial assignment, $\rho : S \rightarrow
            [1,r]$ is the reassignment partial function, $\mu
            \subseteq S \times [1,r] \times S$ is the transition
            relation and $F \subseteq S$ is the set of final states. A
            run of $A$ over an input word $a_1 \ldots a_n \in
            \Sigma^*$ is a sequence $(s_0,\vec{v}_0) \ldots
            (s_n,\vec{v}_n)$ such that $\vec{v}_0 = \vec{u}$ and, for
            all $i \in [1,n]$, exactly one of the following
            holds: \begin{compactitem}
\item if there exists $k\in[1,r]$ such that $a_i=(\vec{v}_{i-1})_k$
  then $\vec{v}_i = \vec{v}_{i-1}$ and $(s_{i-1},k,s_i) \in \mu$, 
\item otherwise $a_i\not\in[\vec{v}_{i-1}]$, $\rho(s_{i-1})$ is
  defined, $(\vec{v}_i)_{\rho(s_{i-1})} = a_i$, for each $k \in
  [1,r] \setminus \set{\rho(s_{i-1})}$, we have $(\vec{v}_i)_k =
  (\vec{v}_{i-1})_k$ and $(s_{i-1},\rho(s_{i-1}),s_i) \in \mu$.
\end{compactitem}
Intuitively, if the input symbol is already stored in some register,
the automaton moves to the next state if, moreover, the transition
relation allows it, otherwise it copies the input to the register
indicated by the reassignment, erasing its the previous value, and
moves according to the transition relation.

The translation of register automata to first order alternating
automata is quite natural, because registers can be encoded as
arguments of predicate atoms. Formally, given a register automaton $R
= \tuple{S,s_0,\vec{u},\rho,\mu,F}$, such that $S =
\set{s_0,\ldots,s_m}$, we define the alternating automaton $\A_R =
\tuple{\set{\alpha},\set{x},Q_R,\iota_R,F_R,\Delta_R}$, where
$\alpha \not\in \Sigma$, $Q_R \isdef \set{q_0,\ldots,q_m}$ and
$\#(q_i) = r$ for all $i \in [m]$, $\iota_R \isdef q_0(\vec{u})$, $F_R
\isdef \set{q_i \mid s_i \in F}$ and, for each transition $(s_i, k,
s_j) \in \mu$, $\Delta_T$ contains the rule:
\[q_i(y_1,\ldots,y_r) \arrow{\alpha(x)}{} y_k=x \wedge q_j(y_1,\ldots,y_r) ~\vee~
\bigwedge_{i=1}^r x \neq y_i \wedge q_j(y_1,\ldots,y_{k-1},x,y_{k+1},
\ldots, y_r)\] Moreover, nothing else is in $\Delta_R$. The connection
between register automata and first order alternating automata is
stated below.

\begin{proposition}\label{prop:register-alternating}
  Given a register automaton $R = \tuple{S,s_0,\vec{u},\rho,\mu,F}$
  over an infinite alphabet $\Sigma$, the first order alternating
  automaton $\A_R = \tuple{\set{\alpha},Q_R,\iota_R,F_R,\Delta_R}$
  recognizes the languge $\lang{\A_R} = \set{ (\alpha,a_1) \ldots
    (\alpha,a_n) \mid a_1 \ldots a_n \in \lang{R} }$. 
\end{proposition}
\proof{ ``$\subseteq$'' Let $w = (\alpha,a_1) \ldots (\alpha,a_n)
  \in \lang{\A_R}$. First, it is easy to show that each execution of
  $\A_R$, that starts in some cube $c \in \cube{\minsem{\iota_R}}$, is
  a linear tree with labels $q_0(\vec{v}_0), \ldots, q_n(\vec{v}_0)$
  such that $\vec{v}_0 = \vec{u}$. Second by induction on $n\geq0$, we
  prove that $\A_R$ has a run as above over $w$ only if $R$ has a run
  $(q_0,\vec{v}_0), \ldots, (q_n,\vec{v}_n)$ over $a_1 \ldots
  a_n$. ``$\supseteq$'' Let $w = a_1 \ldots a_n \in \lang{R}$ and
  $q_0(\vec{v}_0), \ldots, q_n(\vec{v}_0)$ be a run of $R$ over $w$,
  such that $\vec{v}_0 = \vec{u}$. By induction on $n\geq0$, we can
  build an execution of $\A_R$ over $(\alpha,a_1) \ldots
  (\alpha,a_n)$ that is a linear tree with labels $q_0(\vec{v}_0),
  \ldots, q_n(\vec{v}_n)$. \qed}

Consequently, the language inclusion problem ``given register automata
$R_1$ and $R_2$, does $\lang{R_1} \subseteq \lang{R_2}$?'' is reduced
in polynomial time to emptiness problem $\lang{\A_{R_1}} \cap
\lang{\overline{\A_{R_2}}} = \emptyset$, for which
(\S\ref{sec:emptiness}) provides a semi-algorithm. Notice further that
the encoding of register automata as first order alternating automata
uses no transition quantifiers.

\subsection{Predicate Automata}
\label{sec:pa}

The model of \emph{predicate automata} \cite{Farzan15,Farzan16} has
emerged recently as a tool for checking safety and liveness properties
of parameterized concurrent programs, in which there is an unbounded
number of replicated threads that communicate via global
variables. Predicate automata recognize finite sequences of actions
that are pairs $(\sigma,i)$, where $\sigma$ is from a finite set
$\Sigma$ of program statements and $i \in \nat$ ranges over an
unbounded set of thread identifiers. To avoid clutter, we shall view a
pair $(\sigma,i)$ as a data symbol $(\sigma,\nu)$ where $\nu(x)=i$,
for a designated input variable $x$.

Since thread identifiers can only be compared for equality, the data
theory of predicate automata is the first order theory of
equality. Moreover, transition quantifiers are only needed for
checking termination and, generally, liveness properties
\cite{Farzan16}.

However, the execution semantics of predicate automata differs from
that of first order automata with respect to the following detail:
initial configurations and successors of predicate automata are
defined using the entire sets of models of the initial sentence and
transition rules, not just the minimal ones, as in our case. 

Formally, a run of a predicate automaton $P =
\tuple{\Sigma,\set{x},Q,\iota,F,\Delta}$ over a word $(a_1,\nu_1)
\ldots$ $(a_n,\nu_n)$ is a sequence of interpretations $\I_0, \ldots,
\I_n$ such that $\I_0 \in \sem{\iota}$ and for each $i \in [1,n]$,
each $q \in Q$ and each tuple $\tuple{d_1,\ldots,d_{\#(q)}} \in
\I_{i-1}(q)$, we have $\I_i \in \sem{\psi}_\nu$, for each rule
$q(y_1,\ldots,y_{\#(q)}) \arrow{a_i(x)}{} \psi \in \Delta$, where
$\nu=\nu_i[y_1\leftarrow d_1, \ldots, y_{\#(q)} \leftarrow
  d_{\#(q)}]$. The run is accepting if and only if $\I(q)=\emptyset$
for all $q \in Q \setminus F$.

In fact, as shown next, this more simple execution semantics is
equivalent, from the language point of view, with the semantics given
by Definitions \ref{def:execution} and \ref{def:accepting}.  We
believe that the semantics of first order alternating automata based
on minimal models is important for its relation to the textbook
semantics of boolean alternating automata
\cite{ChandraKozenStockmeyer81}.

\begin{proposition}
  Given a predicate automaton $P =
  \tuple{\Sigma,\set{x},Q,\iota,F,\Delta}$, let $\A_P$ be the
  first order alternating automaton that has the same description as
  $P$. Then $\lang{P} = \lang{\A_P}$. 
\end{proposition}
\proof{ ``$\subseteq$'' Let $w = (a_1,\nu_1) \ldots (a_n,\nu_n) \in
  \lang{P}$ be a word and $\I_0, \ldots, \I_n$ be an accepting
  execution of $P$ over $w$. Let $\stamp{\I}{i}_j$ be the
  interpretation that associates each predicate $\stamp{q}{i}$ the set
  $\I_j(q)$, for $i,j \in [n]$. Then one builds, by induction on
  $n\geq0$, an execution $\T$ of $\A_P$ such that $\I_\T \subseteq
  \bigcup_{i=0}^n \stamp{\I}{i}_i$, where $\I_\T$ is the unique
  interpretation associated with $\T$. Since $\I_0, \ldots, \I_n$ is
  accepting, we have $\stamp{\I}{n}_n(\stamp{q}{n}) = \emptyset$, for
  all $q \in Q \setminus F$ and hence $\I_\T(\stamp{q}{n}) =
  \emptyset$, for all $q \in Q \setminus F$ and, consequently $w \in
  \lang{\A_P}$. ``$\supseteq$'' Let $w = (a_1,\nu_1) \ldots
  (a_n,\nu_n) \in \lang{\A_P}$ be a word and $\T$ be an accepting
  execution of $\A_P$ over $w$. We define the sequence of
  interpretations $\I_0, \ldots, \I_n$ as $\I_i(q) =
  \I_\T(\stamp{q}{i})$, for each $i \in [n]$ and each $q \in Q$. By
  induction on $n\geq0$ one shows that $\I_0, \ldots, \I_n$ is an
  execution $P$. Moreover, since $\T$ is accepting, we have $\I_n(q) =
  \I_\T(\stamp{q}{n}) = \emptyset$, for each $q \in Q \setminus F$,
  thus $w \in \lang{P}$. \qed}

As before, this result enables using the
semi-algorithm from \S\ref{sec:emptiness} for checking emptiness of
predicate automata.  We point out that, although quantifier-free
predicate automata with predicates of arity one are decidable for
emptiness \cite{Farzan15}, currently there is no method for checking
emptiness of predicate automata with predicates of arity greater than
one, other than the explicit enumeration of cubes. Moreover, no method
for dealing with emptiness in the presence of transition quantifiers
is known to exist.
\fi

\vspace*{-0.5\baselineskip}
\section{Experimental Results}
\vspace*{-0.5\baselineskip}

We have implemented a version of the IMPACT semi-algorithm
\cite{McMillan06} in a prototype tool called FOADA, which is avaliable
online \cite{foada}. The tool is written in Java and uses the Z3 SMT
solver \cite{z3}, via the JavaSMT interface \cite{javasmt}, for
spuriousness and coverage queries and also for interpolant generation.
The experiments were carried out on a MacOS x64 - 1.3 GHz Intel Core
i5 - 8 GB 1867 MHz LPDDR3 machine.

The experimental results, reported in Table \ref{tab:experiments},
come from several sources, namely predicate automata models (*.pa)
\cite{Farzan15,Farzan16} available online \cite{pa}, timed automata
inclusion problems ({\tt abp.ada}, {\tt train.ada}, {\tt
  rr-crossing.foada}), array logic entailments ({\tt
  array\_rotation.ada}, {\tt array\_simple.ada}, {\tt
  array\_shift.ada}) and hardware circuit verification ({\tt hw1.ada},
{\tt hw2.ada}), initially considered in \cite{IosifRV16}. The {\tt
  train-simpleN.} {\tt foada} and {\tt fischer-mutexN.foada} examples are
parametric verification problems in which one checks inclusions of the
form $\bigcap_{i=1}^N\lang{A_i} \subseteq \lang{B}$, where $A_i$ is
the $i$-th copy of the same template automaton.

The advantage of using FOADA over the INCLUDER \cite{includer} tool
from \cite{IosifRV16} is the possibility of having infinite alphabet
automata with hidden (local) variables, whose values are not visible
in the input. In particular, this is essential for checking inclusion
of timed automata that use internal clocks to control the computation.

\begin{table}[htb]
\vspace*{-\baselineskip}
\begin{center}
{\fontsize{7}{8}\selectfont
\begin{tabular}{||l|c|c|c|c|c||}
\hline
Example & $\len{\A}$ (bytes) & $L(\A)=\emptyset$ ? & Nodes Expanded & Nodes Visited & Time (ms) \\
\hline
incdec.pa & 499 & no & 21 & 17 & 779 \\
\hline
localdec.pa & 678 & no & 49 & 35 & 1814 \\
\hline
ticket.pa & 4250 & no & 229 & 91 & 9543 \\
\hline
count\_thread0.pa & 9767 & no & 154 & 128 & 8553 \\
\hline
count\_thread1.pa & 10925 & no & 766 & 692 & 76771 \\
\hline
local0.pa & 10595 & no & 73 & 27 & 1431 \\
\hline
local1.pa & 11385 & no & 1135 & 858 & 101042 \\
\hline
array\_rotation.ada & 1834 & yes & 9 & 8 & 1543 \\
\hline
array\_simple.ada & 3440 & yes & 11 & 10 & 6787 \\
\hline
array\_shift.ada & 874 & yes & 6 & 5 & 413 \\
\hline
abp.ada & 6909 & no & 52 & 47 & 4788 \\
\hline
train.ada & 1823 & yes & 68 & 67 & 7319 \\
\hline
hw1.ada & 322 & Solver Error & / & / & / \\
\hline
hw2.ada & 674 & yes & 20 & 22 & 4974 \\
\hline
rr-crossing.foada & 1780 & yes & 67 & 67 & 7574 \\
\hline
train-simple1.foada & 5421 & yes & 43 & 44 & 2893 \\
\hline
train-simple2.foada & 10177 & yes & 111 & 113 & 8386 \\
\hline
train-simple3.foada & 15961 & yes & 196 & 200 & 15041 \\
\hline
fischer-mutex2.foada & 3000 & yes & 23 & 23 & 808 \\
\hline
fischer-mutex3.foada & 4452 & yes & 33 & 33 & 1154 \\
\hline
\end{tabular}
}
\caption{Experiments with First Order Alternating Automata}
\label{tab:experiments}
\end{center}
\vspace*{-2\baselineskip}
\end{table}

\bibliographystyle{abbrv} \bibliography{refs}

\end{document}